\documentclass[useAMS,usenatbib]{mn2e}
\usepackage{graphicx}
\usepackage{float}
\title[The origin of gas in the ENLR]{The origin of gas in the Extended Narrow Line Region of nearby Seyfert galaxies.I. NGC 7212\thanks{This paper includes data gathered with the 6.5 meter Magellan Telescopes located at Las Campanas Observatory, Chile, and it is partly based on observations made with the 6-m telescope of the SAO RAS operated under the financial support of the Ministry
of Science of the Russian Federation (Registration Number 01-43).}}

\author[V. Cracco et al.]{V. Cracco$^{1}$\thanks{E-mail:valentina.cracco@unipd.it},
S. Ciroi$^{1}$, 
F. Di Mille$^{2,1}$, L. Vaona$^{1}$, A. Frassati$^{1}$, A.A. Smirnova$^{3}$, 
\newauthor G. La Mura$^{1}$, A.V. Moiseev$^{3}$, P. Rafanelli$^{1}$  \\
$^{1}$Department of Astronomy, Padova University, Vicolo dell'Osservatorio 3, Padova, 35122, Italy\\
$^{2}$Australian Astronomical Observatory-Carnegie Observatories  Colina El Pino, Casilla 601 La Serena, Chile\\
$^{3}$Special Astrophysical Observatory, Nizhnij Arkhyz, 369167 Russia
}
\begin{document}

\date{}

\pagerange{\pageref{firstpage}--\pageref{lastpage}} \pubyear{}

\maketitle

\label{firstpage}

\begin{abstract}

The Extended Narrow Line Region (ENLR) of Active Galactic Nuclei (AGN) is a region of highly ionized gas with a size of few up to 15--20 kpc. When it shows a conical or bi-conical shape with the apexes pointing towards the active nucleus, this region is also called ionization cones. The ionization cones are an evidence of the Unified Model that predicts an anisotropic escape of ionizing photons from the nucleus confined to a cone by a dusty torus.
Many details about the complex structure of the ENLR still remain unveiled, as for example the origin of the ionized gas. 
Here we present new results of a study of the physical and kinematical properties of the circumnuclear gas in the nearby Seyfert 2 galaxy NGC 7212. Medium and high resolution integral field spectra and broad-band photometric data were collected and analysed in the frame of an observational campaign of nearby Seyfert galaxies, aiming to handle the complicated issue of the origin of the gas in the ENLR. 
This work is based on: (i) analysis of gas physical properties (density, temperature and metallicity), (ii) analysis of emission line ratios, and (iii) study of kinematics of gas and stars. 
By reconstructing the [O\,{\sc iii}]/H$\beta$ ionization map, we pointed out for the first time the presence of an ionization cone extended up to about 6 kpc, made by a large amount of low metallicity gas, kinematically disturbed and decoupled from stars,  whose highly ionized component shows radial motions at multiple velocities proved by the complex profiles of the specral lines.
Since NGC 7212 is a strongly interacting triple galaxy system, the gravitational effects are likely to be at the origin of the ENLR in this Seyfert galaxy.
\end{abstract}

\begin{keywords}
galaxies: Seyfert -- galaxies: individual: NGC 7212 --  techniques: spectroscopic -- line: profiles 
\end{keywords}

\section{Introduction}
Ionization cones are an evidence of the validity of the Unified Model \citep{1993ARA&A..31..473A}, which postulates an anisotropic escape of photons from the Active Galactic Nuclei (AGN) confined to a cone by a dusty torus. Up to now, a bright Extended Narrow Line Region (ENLR) was found in only 25 galaxies. Many details about the complex structure of the ENLR still remain unveiled. 
The small number of ionization cones could be related to the still open problem of the origin of the ionized gas in the ENLR.
In fact, the ENLR gas could be part of the interstellar medium (ISM) of the host galaxy, photoionized by the active nucleus, or material ejected by the nucleus in strong interaction with a radio-jet. Otherwise, the gas could be acquired from the intergalactic medium, if the galaxy is in a dense environment or by means of gravitational interactions, in case of minor merger events. 
This issue is strongly related to the mechanisms of AGN feeding. 

Within this frame, we are carrying out an observational program taking advantage of the integral-field spectroscopic technique to investigate the physical and kinematical properties of the ENLR in few nearby Seyfert 2 galaxies (z $<$ 0.03) with possible ionization cones or at least extended [O\,{\sc iii}] emission.
Here we present new results of the first studied object, NGC 7212.

NGC 7212 (z=0.0266) is a Seyfert 2 galaxy in an interacting system of three galaxies \citep{1981PASP...93..560W}. \citet{1990A&A...238...15D} by means of optical longslit spectroscopy in the H$\beta$ + [O\,{\sc iii}] spectral range, found ionized gas extending for $\sim17$ arcsec along PA = 37\degr\ and 127\degr, and a high excitation value R=I([O\,{\sc iii}])$\lambda$ 5007+4959/I(H$\beta$)=19 in the nucleus, ranging from 5 to 28 in the ionized emission. \citet{1995ApJ...440..578T} detected a jet-like highly ionized feature extended up to 10 arcsec from the nucleus at PA = 170\degr\ in ground-based [O\,{\sc iii}] and H$\alpha$ images. This feature is exactly parallel to the axis of the small scale double radio source (0.7 arcsec separation) published by \citet{1998ApJ...502..199F}, who did not find a clear evidence of an ionization cone, since the Northern side of the galaxy is partially obscured by dust lanes. However, they suggested that the distribution of excitation towards the South is consistent with a ragged ionization cone.
\citet{1998A&AS..132..197K} published the $B-I$ map of NGC 7212 showing a very blue and fan-shaped emission region extending from the nucleus towards South (PA = 165\degr) with total size of 2.3 arcsec, and a dust lane situated on the other side of the nucleus, at 3.7 arcsec and PA = 280\degr. 
{\em Hubble Space Telescope (HST)} [O\,{\sc iii}] images by \citet{2003ApJS..148..327S}  showed that the [O\,{\sc iii}] emission is extended up to $\sim$3 arcsec from the nucleus along PA = 170\degr\ with dimensions $2.1 \times 4.8$ arcsec. The emission is diffuse and composed of several individual knots Northwards and Southwards the nucleus.
\citet{2003MNRAS.339..772R} found that the nuclear stellar component of NGC 7212 is dominated by 10-Gyr metal rich (solar or above solar metallicity) stellar population, with a contribution of 15 per cent of the total flux from the 1-Gyr component and 15 per cent from the 3-Myr component or from a featureless continuum.
\citet{2006A&A...456..953B} confirmed the [O\,{\sc iii}] emission along PA = 170\degr, but extended up to 12 arcsec from the nucleus, i.e. four times larger than the extension seen in the {\em HST} image in the same direction, but smaller than the maximum extent reported by \citet{1990A&A...238...15D}, who observed at different position angles.
The reddening corrected excitation value in the central spectrum is $\rm R\sim$16 and varies between 6 and 17 in the central 24-arcsec-wide region. 
The reddening in the centre is rather low (E(B$-$V)$=0.33\pm0.01$ mag) and decreases to a value of $\sim$0.07 mag at 1 arcsec North-Westerly of the nucleus. On both sides of this region, it increases and reaches its maximum value at 4 arcsec South-Easterly and 7 arcsec North-Westerly of the photometric centre ($\Delta$E(B$-$V)$\sim$1 mag). These maxima may be attributed to dust lanes seen in the continuum image by \citet{1998ApJ...502..199F}.

\section[]{Observations and data reduction}
NGC 7212 was observed at the Russian 6-m telescope of the Special Astrophysical Observatory (SAO RAS) with the Multi Pupil Fiber Spectrograph \citep[hereafter MPFS,][]{2001sdcm.conf..103A}. This integral field unit takes simultaneous spectra from 256 spatial elements, the field of view (FoV) is 16 square arcsec with a spatial sampling of 1 arcsec. 
Spectra of NGC 7212 were obtained in September 2005 in the range 3800--7300 \AA\ with a 600 lines mm$^{-1}$ grating, and in July 2007 in the range 4000--7300 \AA\ with a 1200 lines mm$^{-1}$ grating. The data have dispersion of 1.5 \AA\ px$^{-1}$ and 0.8 \AA\ px$^{-1}$, and instrumental resolution of $\sim$6 \AA\ and $\sim$3 \AA, respectively. 
\begin{figure}
\centering
\includegraphics[width=8.5cm]{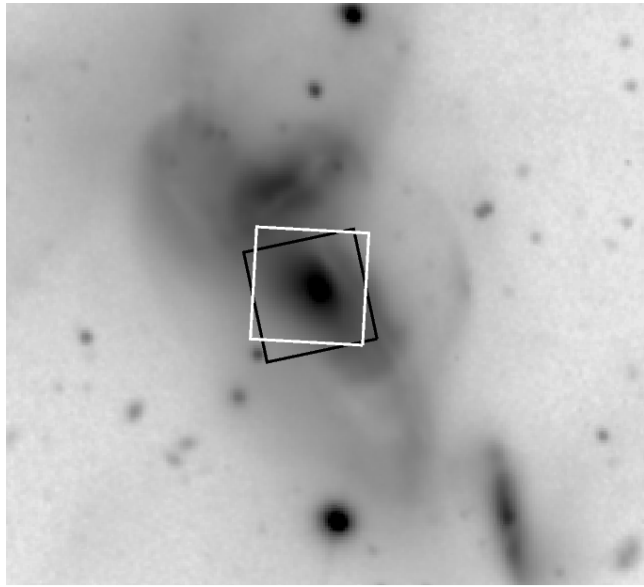}
\caption{$R$-band image of NGC 7212 obtained at the SAO 6-m telescope with SCORPIO. The white (black) square is the MPFS FoV for the low (high) resolution data. The FoV is 16$\times$16 square arcsec, corresponding to 8.2$\times$8.2 kpc$^{2}$ (scale = 0.513 kpc arcsec$^{-1}$, $\rm H_0=75$ km s$^{-1}$ Mpc$^{-1}$). North is up and East is on the left.}
\label{ngc7212}
\end{figure}
NGC 7212 was also observed with the high resolution (R$\sim$8000) MagE echelle spectrograph at the Magellan Telescopes of the Las Campanas Observatory (Chile), with the slit oriented along the ionization cones (PA$\sim$170\degr). From HST-MAST archive a WFPC2 image in the F606W broad band filter was extracted. 
Broad band $BVR$ images were obtained in September 2008 at the SAO 6-m telescope with SCORPIO \citep{2005AstL...31..194A} (Fig. \ref{ngc7212}). 
All the observational and archival data are listed in Table \ref{tabella1}, where $\Delta \lambda$ is the spectral resolution in \AA\ and $\delta \lambda$ is the dispersion in \AA\ px$^{-1}$. 
\begin{table}
\caption{The observational and archival data}
\tiny
  \begin{tabular}{cccccc}
\hline
Date & Instrumental & Spectral range & $\Delta \lambda$ & $\delta \lambda$ & T$\rm _{exp}$ \\  
     &            &  (\AA) & (\AA) & (\AA\ px$^{-1}$) & (s)\\
\hline
Sep 2005 &MPFS & 3780--5800  & 6  & 1.46  & 4800\\ 
Sep 2005 &MPFS & 4340--7310  & 6  & 1.47  & 5400\\
Jul 2007 &MPFS & 4140--5650  & 3  & 0.76  & 4800\\ 
Jul 2007 &MPFS & 5765--7260  & 3  & 0.76  & 1800\\
May 2010 &MagE & 3600--7000  & 0.85  & -- & 7200\\
Sep 2008 &SCORPIO & $B$ & -- & -- & 600\\
Sep 2008 &SCORPIO & $V$ & -- & -- & 660\\
Sep 2008 &SCORPIO & $R$ & -- & -- & 510\\
Sep 1994 &WFPC2   & F606W & -- & -- & 500\\
\hline
    \end{tabular}
\label{tabella1}
\end{table}

The MPFS data were reduced using {\sc p}3d \citep{2010A&A...515A..35S} which is a general data-reduction software written in {\sc idl} language and developed to work with fibre-fed integral-field units (IFU) of any integral-field spectrograph. Up to now, it has been configured and tested with PMAS, VIRUS-P, SPIRAL and MPFS. {\sc p}3d makes the reduction steps up to wavelength calibration automatically, and allows to interactively inspect and optimize parameters when required. 
It also provides graphical tools to check the raw data and the output of the different tasks. 
In order to perform the flux calibration, we observed a spectrophotometric standard star for each run. Its spectra were extracted, flat fielded and wavelength calibrated with {\sc p}3d, then sky subtracted and summed together, to collect the total flux, with {\sc iraf}.
The usual {\sc iraf} tasks for the flux calibration were applied to the data. 
The echelle spectrum was reduced with {\sc iraf} following the standard procedures (bias subtraction, flat-field correction and wavelength calibration). 
No other calibrations or corrections were applied. 
Before reducing this spectrum, the orders were extracted using {\sc apall} with the option \texttt{strip}. 

After the reduction procedures, the IFU spectra were corrected for Galactic reddening, using the {\sc iraf} task {\sc deredden} and the value of \emph{V}-band absorption (A$_V=0.238$) derived from NED (NASA/IPAC Extragalactic Database), then we corrected for telluric absorptions. 
In fact, at $\rm z=0.0266$ the [S\,{\sc ii}] doublet falls in the range 6880--6930 \AA, and therefore it is only partially absorbed by the atmospheric B band (6860--6890 \AA). 
We used two {\sc iraf} scripts, {\sc atmo} and {\sc rmat}, written by our group. {\sc atmo} fits the continuum of the flux-calibrated standard star and divides the fitting by the stellar spectrum. Then, all the intensity values at a $\lambda$ shorter than 6700 \AA\ are put equal to 1. Finally, the corrected spectrum is obtained using the task {\sc rmat} which simply multiplies the observed spectrum by the output of {\sc atmo}. 
We also applied a correction for atmospheric refraction. 
We used two {\sc iraf} scripts written by us, named {\sc acorr} and {\sc darcspec}. The first task determines the correction to apply for atmospheric refraction, the second one applies this correction. In particular, {\sc acorr} creates images  of the emission for 20--30 intervals centred at different $\lambda$ along the dispersion direction, then it calculates the centroid of each image using {\sc imcentroid} and allows to interactively fit with a straight line the relations $\lambda$ vs. \emph{x} and $\lambda$ vs. \emph{y}, using {\sc nfit1d}. {\sc darcspec} uses the \emph{x}$(\lambda)$ and \emph{y}$(\lambda)$ functions and for each $\lambda$ along all the spectral range, makes an image, applies the needed shift (taking as a reference the centroid at $\lambda$ = 5500 \AA) and then transforms the image in a table with three columns: \emph{x}, \emph{y}, $\lambda$. From these output tables the spectrum corrected for atmospheric refraction is reconstructed. 

In order to accurately measure the emission lines, in particular hydrogen and helium Balmer lines, the underlying stellar component was subtracted from the spectra. 
We applied the {\sc starlight} software \citep{2007MNRAS.375L..16C,2005MNRAS.358..363C} to fit the galactic stellar component. 
Before being analysed with {\sc starlight}, the spectra were shifted to rest-frame wavelengths with {\sc newredshift} and their dispersion was modified to a value of 1 \AA\ px$^{-1}$ with {\sc dispcor}. 
We used 45 synthetic spectra by combining 15 ages (from $10^6$ yr up to $13\times10^9$ yr) with 3 metallicities (Z = 0.004, 0.02 and 0.05 Z$_{\odot}$), and the \citet*{1989ApJ...345..245C} (hereafter CCM) extinction function. The synthetic spectrum of the stellar contribution obtained from {\sc starlight} for each fibre was subtracted from the observed one to obtain a pure emission line spectrum.
These procedures were performed only for low resolution spectra, because high resolution data were used only to study the kinematics. Moreover, in high resolution spectra the Balmer absorption lines could be neglected because of the low S/N ratio (less than 5) of the continuum.

For the analysis of IFU data, we used {\sc pan} (Peak ANalysis), an {\sc idl} software that fits emission lines with a graphical interface, based on Craig Markwardts {\sc mpfit}. We applied a single Gaussian function for all the emission lines except H$\alpha$ + [N\,{\sc ii}] and [S\,{\sc ii}] for which we used custom functions which allowed to fit all the lines simultaneously. 
{\sc pan} is useful for integral field data since it can read multiple spectra, and fit the initial guess to all the spectra automatically. The spectrum, fitting and residuals can be reviewed, being displayed on-screen. 
The initial parameters (amplitude, position and width) can be specified interactively and  constrained to ensure a reasonable result: 
it is possible to fix to the same value the FWHM of the doublets, like [N\,{\sc ii}] and [S\,{\sc ii}]. 
For the high resolution data of NGC 7212 we used the {\sc midas} package {\sc xa}lice because it was necessary to fit the [O\,{\sc iii}] emission lines with two components, a broad and a narrow one. 

\begin{figure}
\centering
\includegraphics[width=4cm]{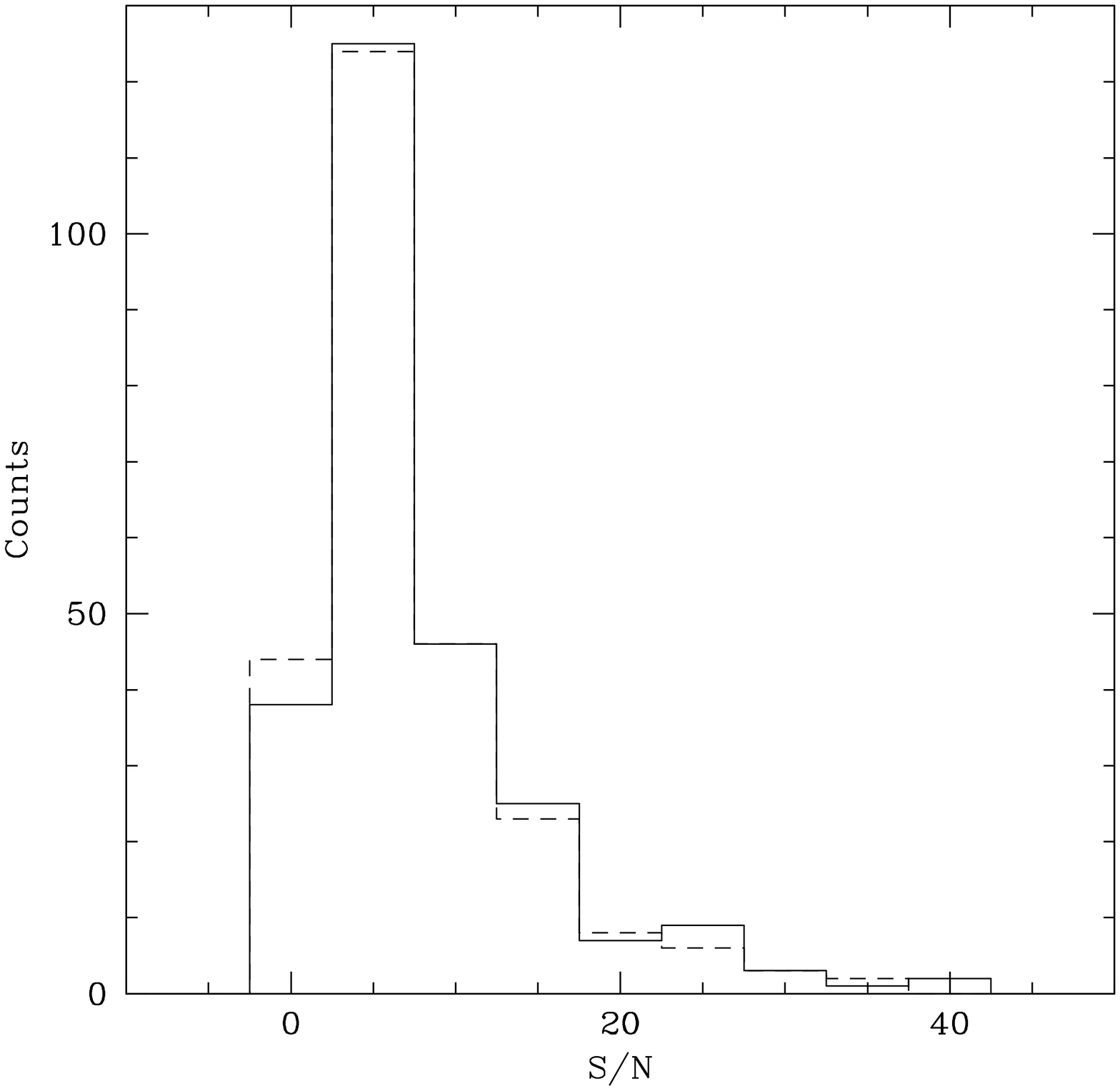}
\includegraphics[width=4cm]{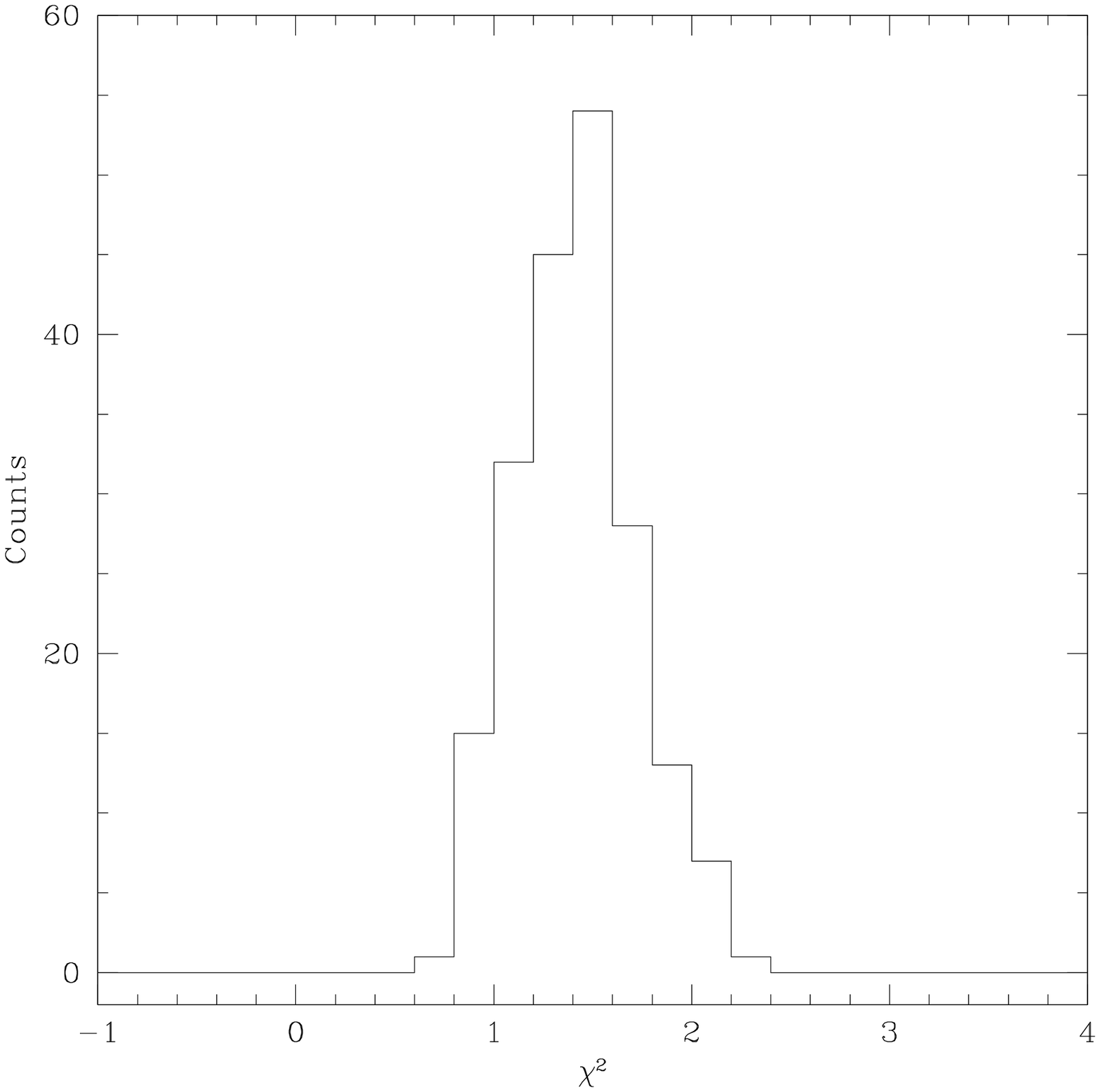}
\caption{Left: the S/N of the IFU low resolution data measured at 5500 \AA\ (solid line) and at 7000 \AA\ (dashed line). Right: the histogram of the $\chi^2$, evaluating the quality of the {\sc starlight} fitting.}
\label{SN}
\end{figure}

The reliability of our measurements was verified by calculating the signal to noise  ratio (S/N) of the continuum for MPFS data (see Fig. \ref{SN}). These data have small values of S/N: the median values for low resolution data are $\sim$5.6 and $\sim$5.4, measured at rest-frame 5500 \AA\ and 7000 \AA. 
The $\chi^2$ evaluates the quality of the {\sc starlight} fitting: 
it is peaked around 1.6 (see Fig. \ref{SN}), instead of 1 probably because the S/N ratio of the analysed spectra is not sufficiently high. 
We calculated the flux errors, assuming that the determination of the position of the continuum level gives the main contribution to the errors when measuring the emission line fluxes.

\section[]{Physical properties of gas and stars}
\begin{figure}
\centering
 \includegraphics[height=4cm]{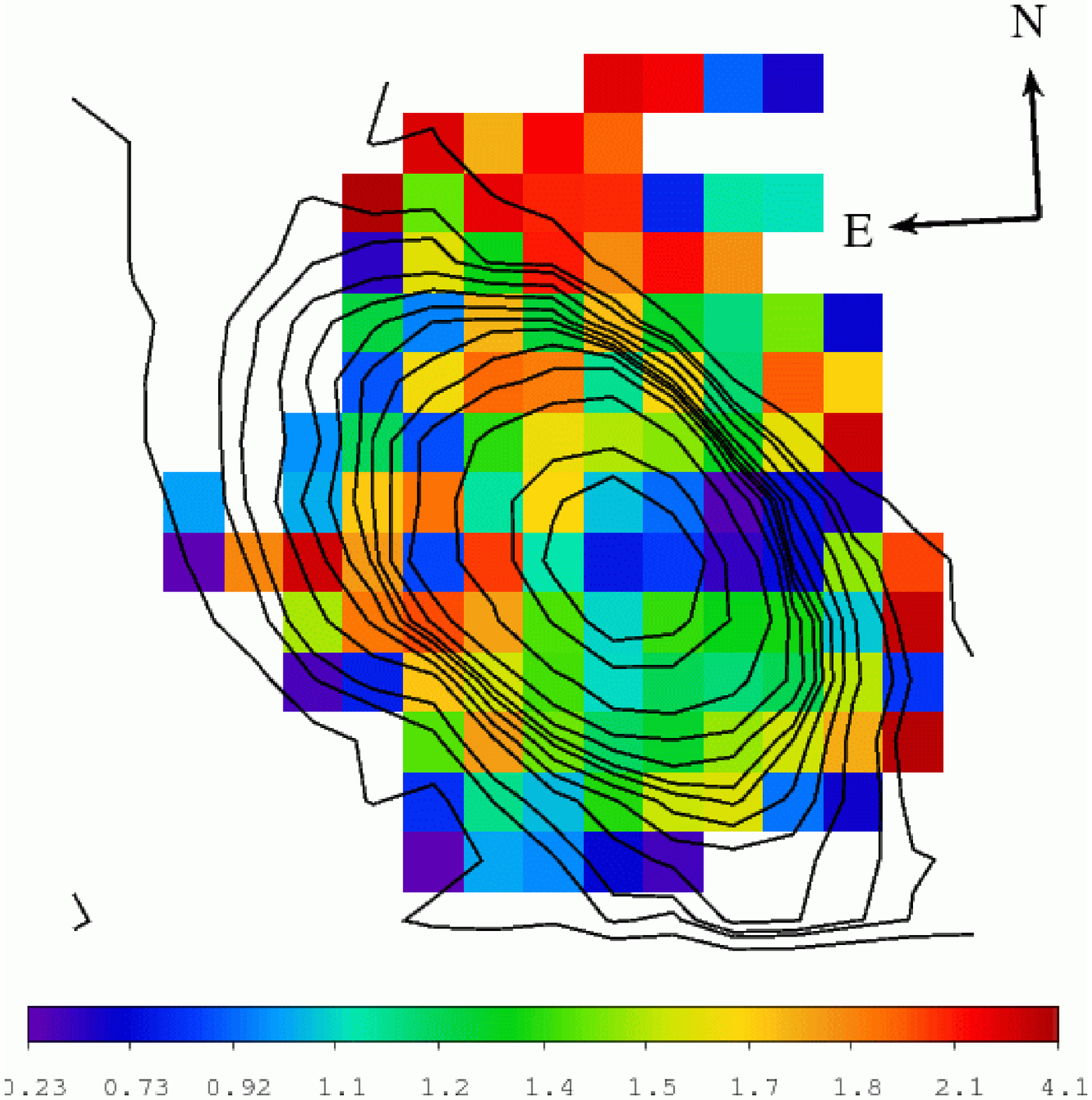}
 \includegraphics[height=4cm]{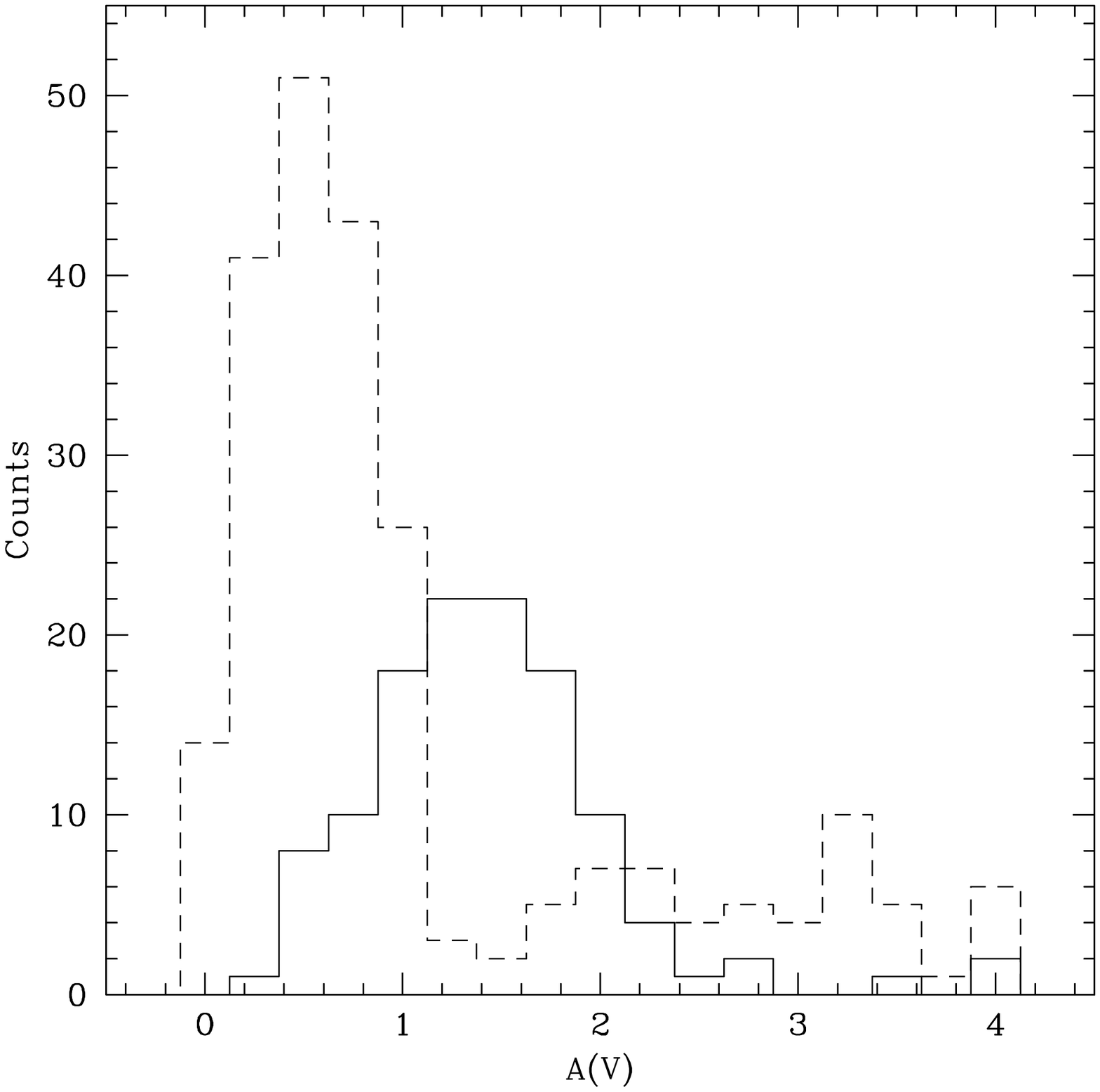}
\caption{Left: A($V$) map, obtained by applying the CCM extinction law to the ratio of the intensity of H$\alpha$\ and H$\beta$, with the stellar continuum contours overlaid (in black). Right: The histogram of the A($V$) values obtained for gas (solid line) and stars (dashed line).}
\label{Av}
\end{figure}

All the measured fluxes were corrected for the internal reddening, using the Balmer decrement and assuming the theoretical value of the intensity ratio between H$\alpha$ and H$\beta$ equal to 2.86.
The visual absorption, A($V$), was estimated by applying the CCM extinction law. 
Once we obtained A($V$), we corrected all the observed fluxes. 
In order to study the spatial distribution of the extinction, we reconstructed the A($V$) map (see Fig. \ref{Av}), which is clearly limited to the region where the H$\beta$ emission line can be measured. 
In the nucleus the reddening is low, A($V$) = 0.78 mag, higher values, A($V$) = 1.6--2 mag, are found at about 7 arcsec North-West and at about 4 arcsec East of the nucleus, in agreement with the reddening obtained by \citet{2006A&A...456..953B}, E(B$-$V) = 0.33 mag in the centre and E$(B-V)\simeq0.8$ mag in the northwestern and southeastern regions (see their fig. 3, upper-right panel).  
We estimated the extinction from the stellar spectra by means of {\sc starlight}, which also gives as output the A($V$) values for each spectrum. By comparing the A($V$) obtained from gas and stars, we can see that the distributions are peaked at different values, larger for the gas: the median value for gas is 1.38 mag and for stars is 0.68 mag. 
If the ENLR gas were dust-free, then the A($V$) measured by stars and gas would be the same. 
The larger extinction observed in the gas component suggests that there is also a fraction of dust associated to the ENLR.

\begin{figure}
 \centering
\includegraphics[height=4cm]{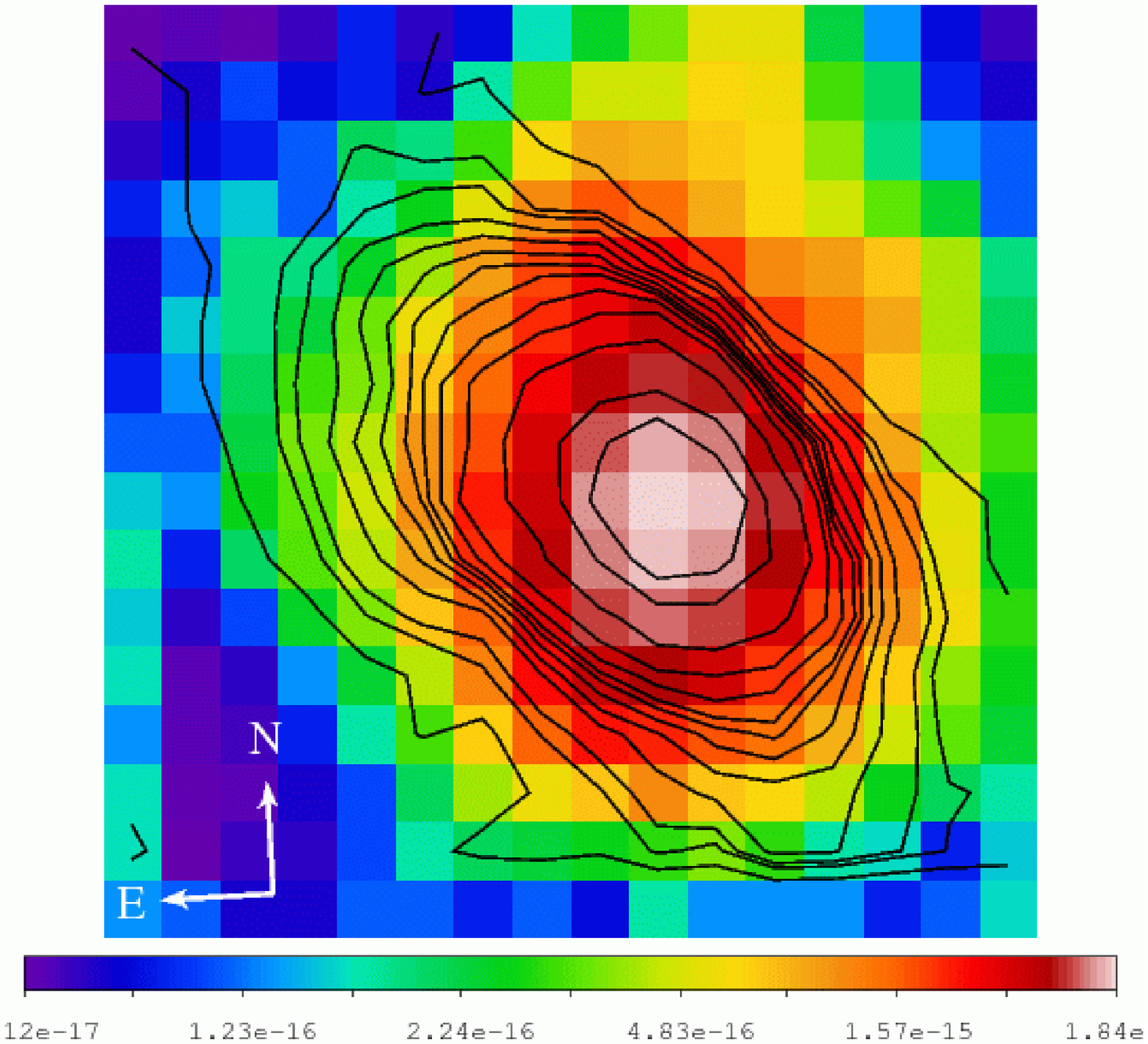}
\includegraphics[height=4cm]{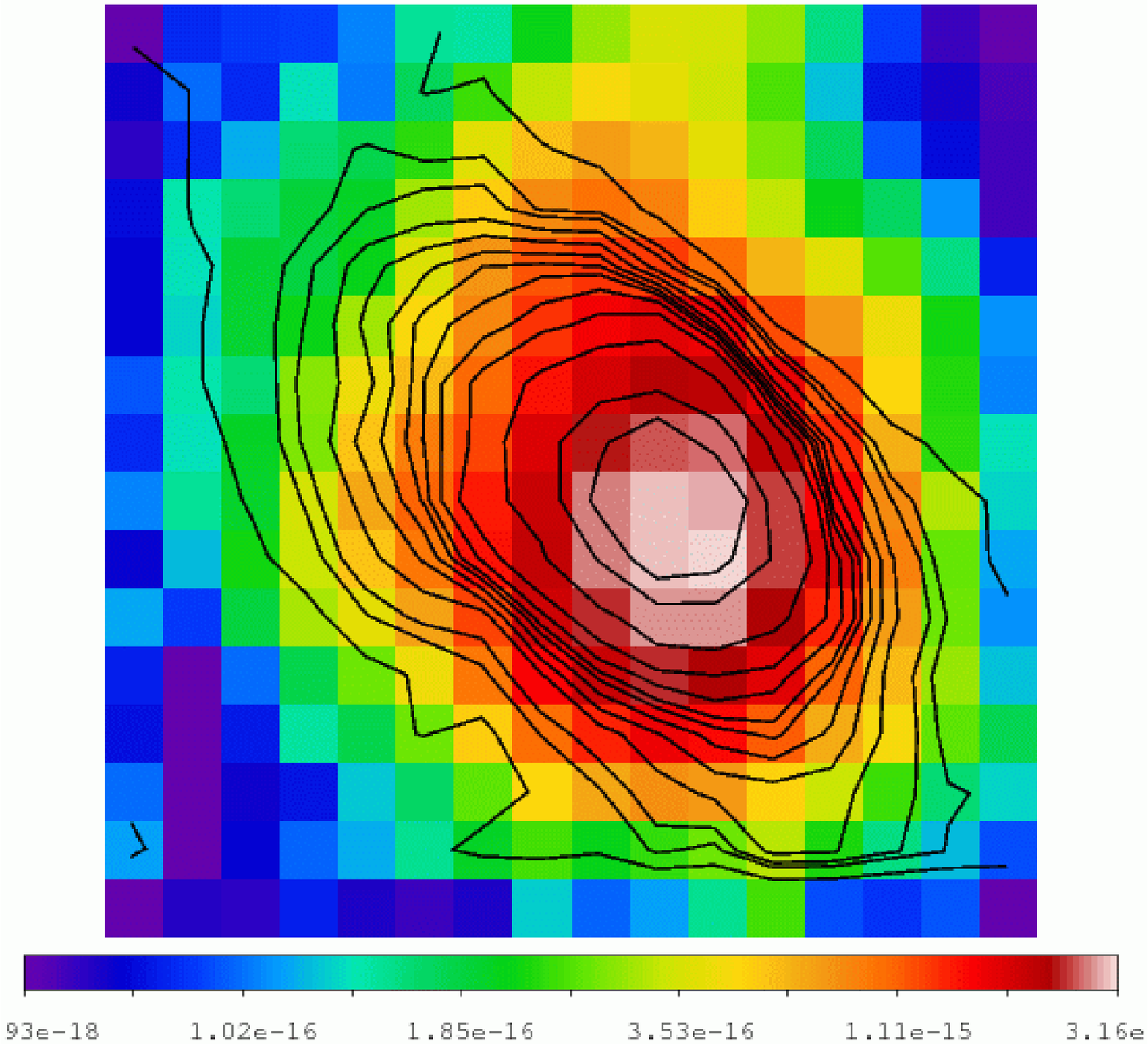}
\caption{Left: the [O\,{\sc iii}] emission line map with the overlaid contours of the stellar continuum emission at 5500 \AA\ in black. Right: the same for H$\alpha$. The gas emission is oriented along PA = 0\degr\ and the stellar continuum is oriented along PA = 45\degr. North is up and East is on the left. }
\label{oiii_ha_cont5500}
\end{figure}

\begin{figure}
\centering
\includegraphics[height=4.5cm]{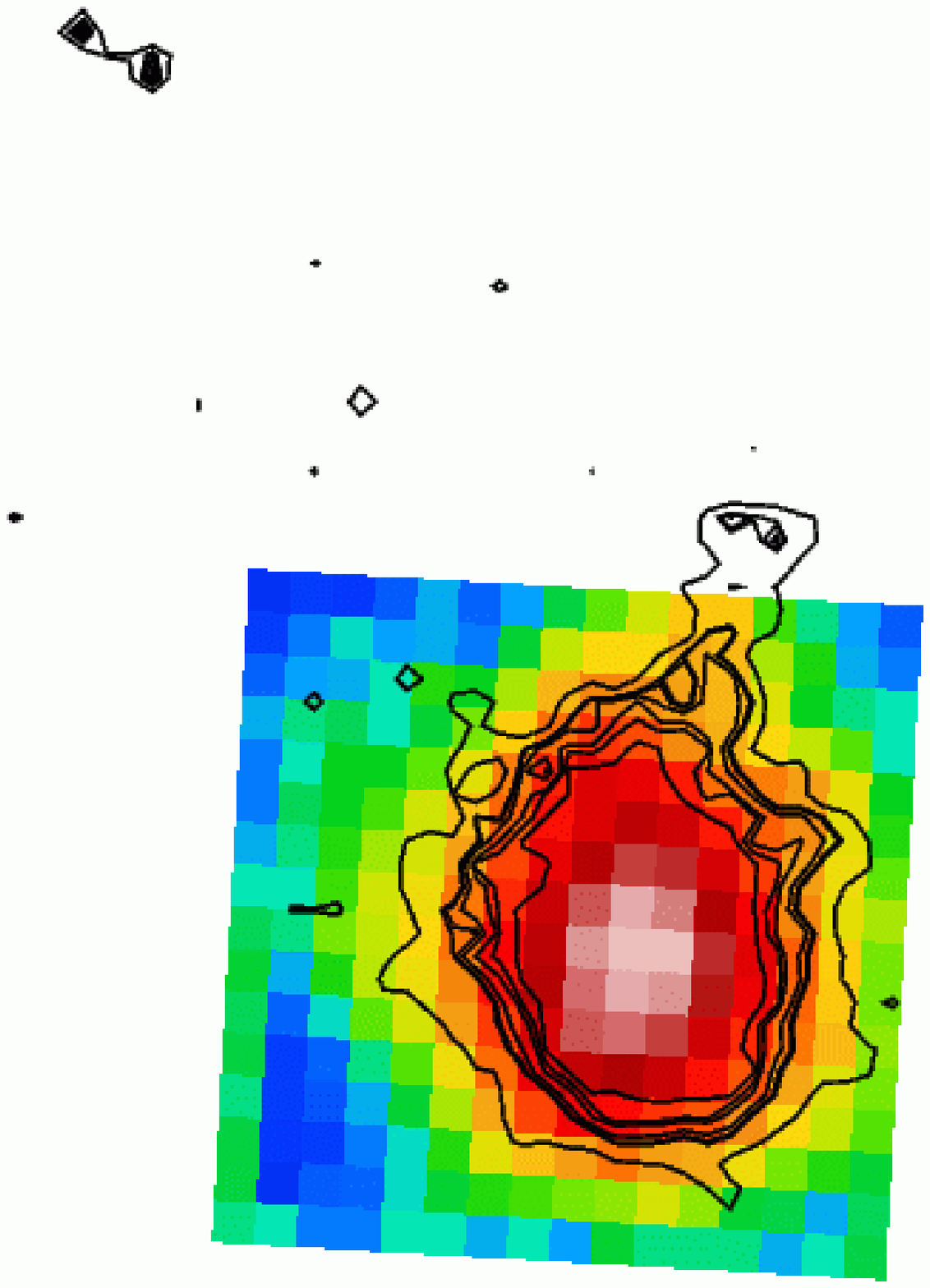}
\includegraphics[height=4.5cm]{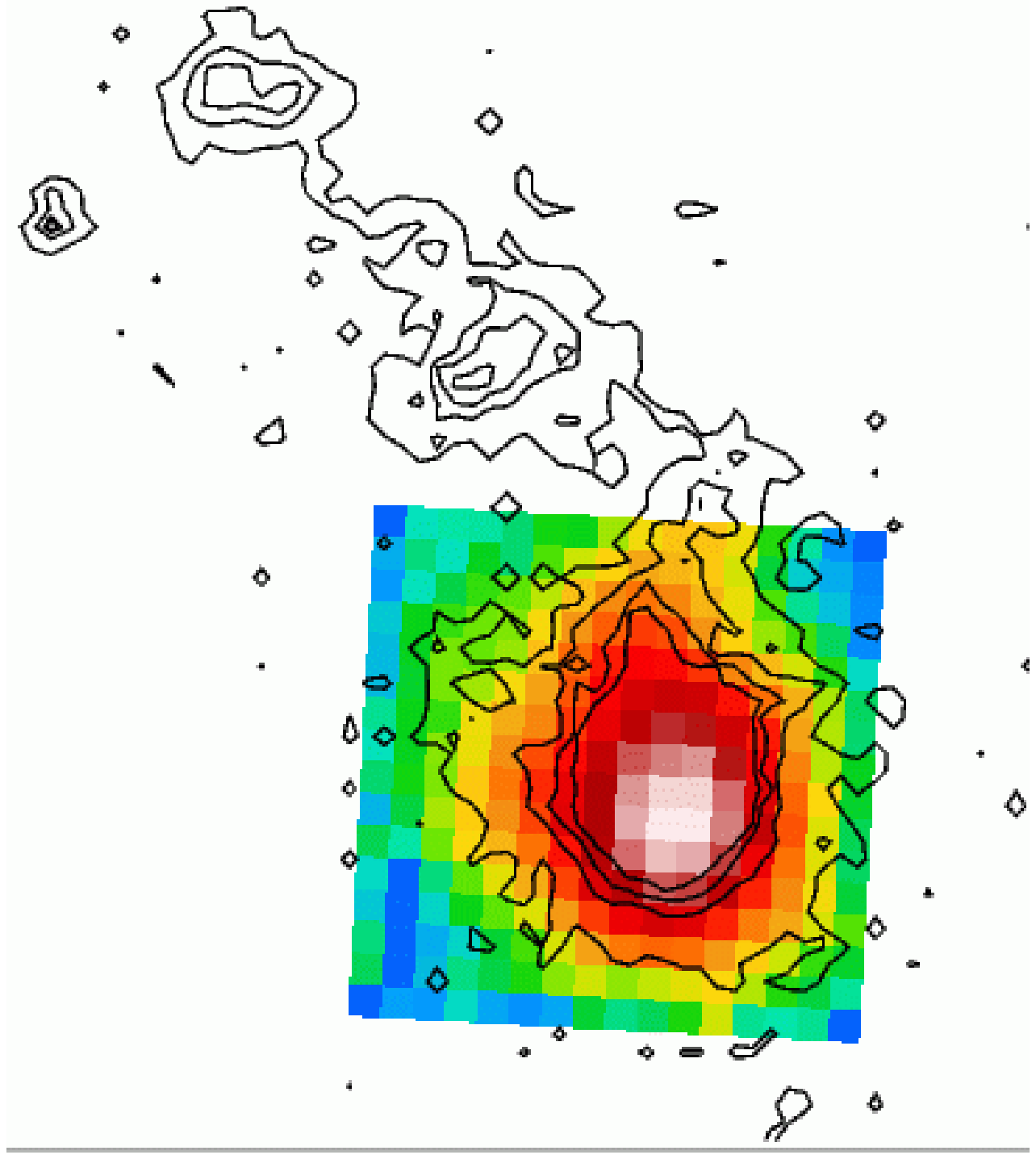}
\caption{Left: 2D map of the flux distributions for [O\,{\sc iii}] with the \citet{1995ApJ...440..578T} contours overlaid in black. Right: the same for H$\alpha$. North is up and East is on the left.}
\label{o3ha_cont_tran95}
\end{figure}

\begin{figure}
\centering
\includegraphics[width=9cm]{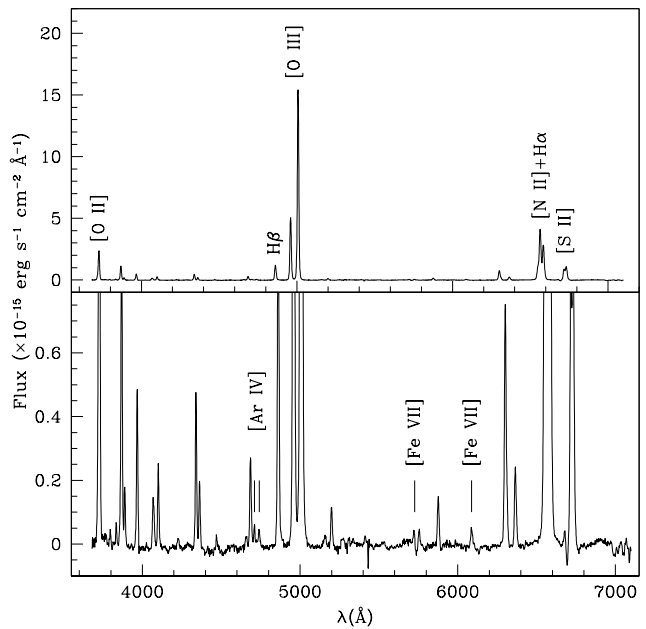}
\caption{The nuclear spectrum of NGC 7212. In the upper panel the brightest emission lines are clearly visible, in the bottom panel the Argon and Iron emission lines are shown.}
\label{nucl_spettro}
\end{figure}

Up to 30 emission lines were detected and measured at least in the central regions, within a radius of about 4 arcsec. 
The maps of the brightest emission lines ([O\,{\sc ii}]\ $\lambda$3727, [O\,{\sc iii}] $\lambda$4959, [O\,{\sc iii}] $\lambda$5007, [O\,{\sc i}] $\lambda$6300, [N\,{\sc ii}] $\lambda$6548, H$\alpha$ $\lambda$6563, [N\,{\sc ii}] $\lambda$6584, [S\,{\sc ii}] $\lambda$6716, [S\,{\sc ii}] $\lambda$6731) show an elongated shape: the ionization gas is extended up to 15 arcsec (corresponding to $\sim$7.7 kpc, scale = 0.513 kpc arcsec$^{-1}$) with a PA $\simeq$ 0\degr.  
A similar size was detected by \citet{1990A&A...238...15D} with longslit spectroscopy. They found ionized gas extending for about 17 arcsec at PA = 37\degr\ and PA = 127\degr, corresponding to the minor and major axis of the galaxy, respectively. We verified that this emission is within our FoV.
\citet{2006A&A...456..953B} found ionized gas extending out to 12 arcsec along a PA $\simeq$ 170\degr, as well.
We compared our emission lines maps of [O\,{\sc iii}], H$\alpha$ and [N\,{\sc ii}] with the images published by \citet{1995ApJ...440..578T}. Fig. \ref{o3ha_cont_tran95} shows a good match between the contours of these images and our two-dimensional (2D) maps.
The emission is not oriented as the stellar continuum emission which has the major axis at PA = 45\degr, in Fig. \ref{oiii_ha_cont5500} the [O\,{\sc iii}]\ and H$\alpha$ emission line maps are shown with the contours of the stellar continuum emission at 5500 \AA\ overlaid.
[O\,{\sc iii}] $\lambda$4363 and He\,{\sc ii} $\lambda$4686 maps are clearly elongated as well with PA $\simeq$ 0\degr\ and extended up to 9--10 arcsec (corresponding to $\sim$ 4.5--5 kpc).  Fainter emission lines are visible not only in the very inner parts, where the influence of the ionization source is more important, but also at 3--6 arcsec (1.5--3 kpc) from the nucleus.
The coronal emission lines of iron with different ionization degree were detectable as well and were measured, but only in the central spectra, probably because here both the ionization and the S/N are sufficiently high. We detected [Fe\,{\sc iii}]\ $\lambda$4658, [Fe\,{\sc v}]\ $\lambda$4228, [Fe\,{\sc vii}]\ $\lambda$5158 and [Fe\,{\sc vii}]\ $\lambda$5721 and [Fe\,{\sc vii}]\ $\lambda$6087 (Fig. \ref{nucl_spettro}).

\begin{figure}
 \includegraphics[width=4cm]{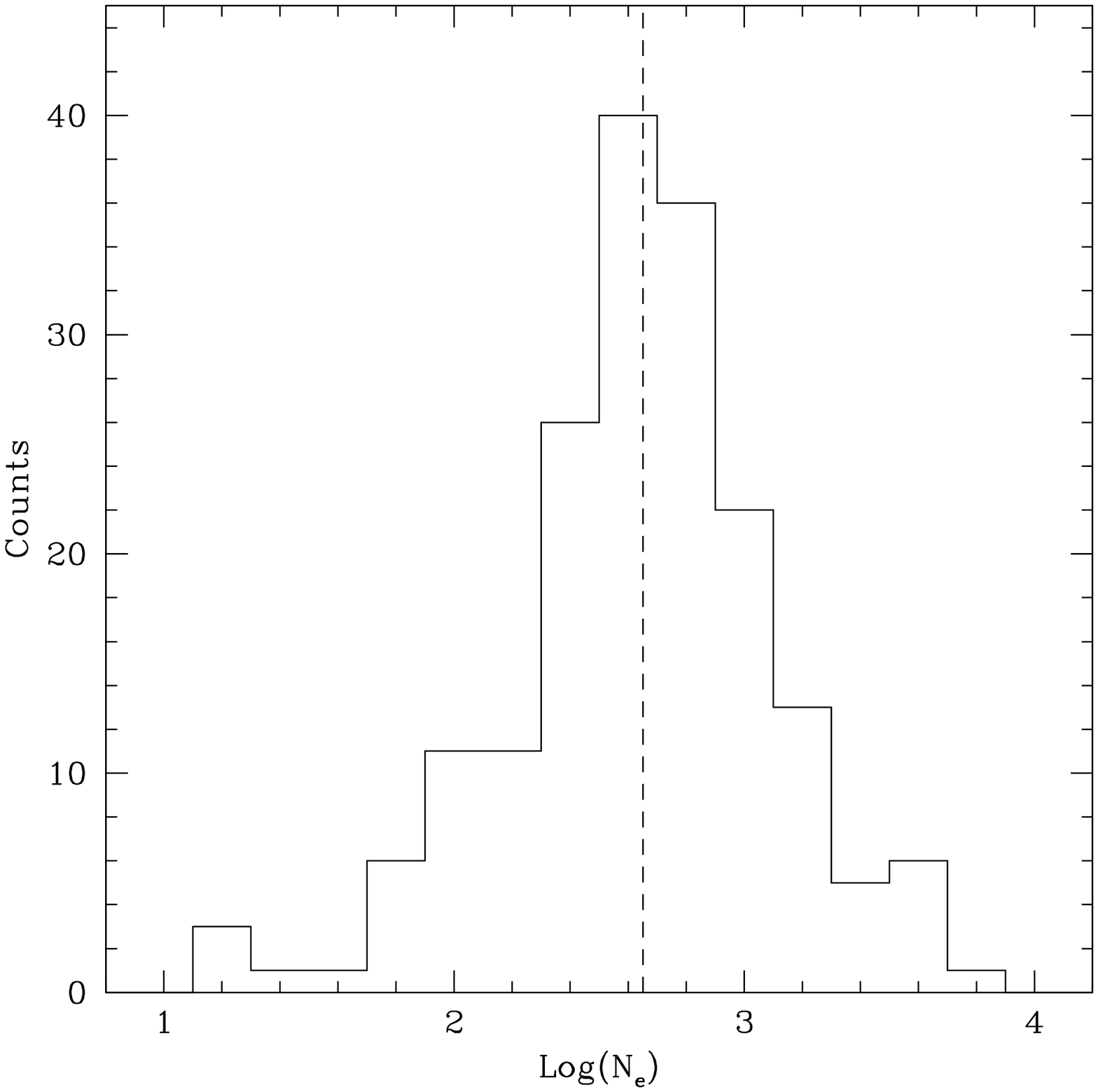}
 \includegraphics[width=4cm]{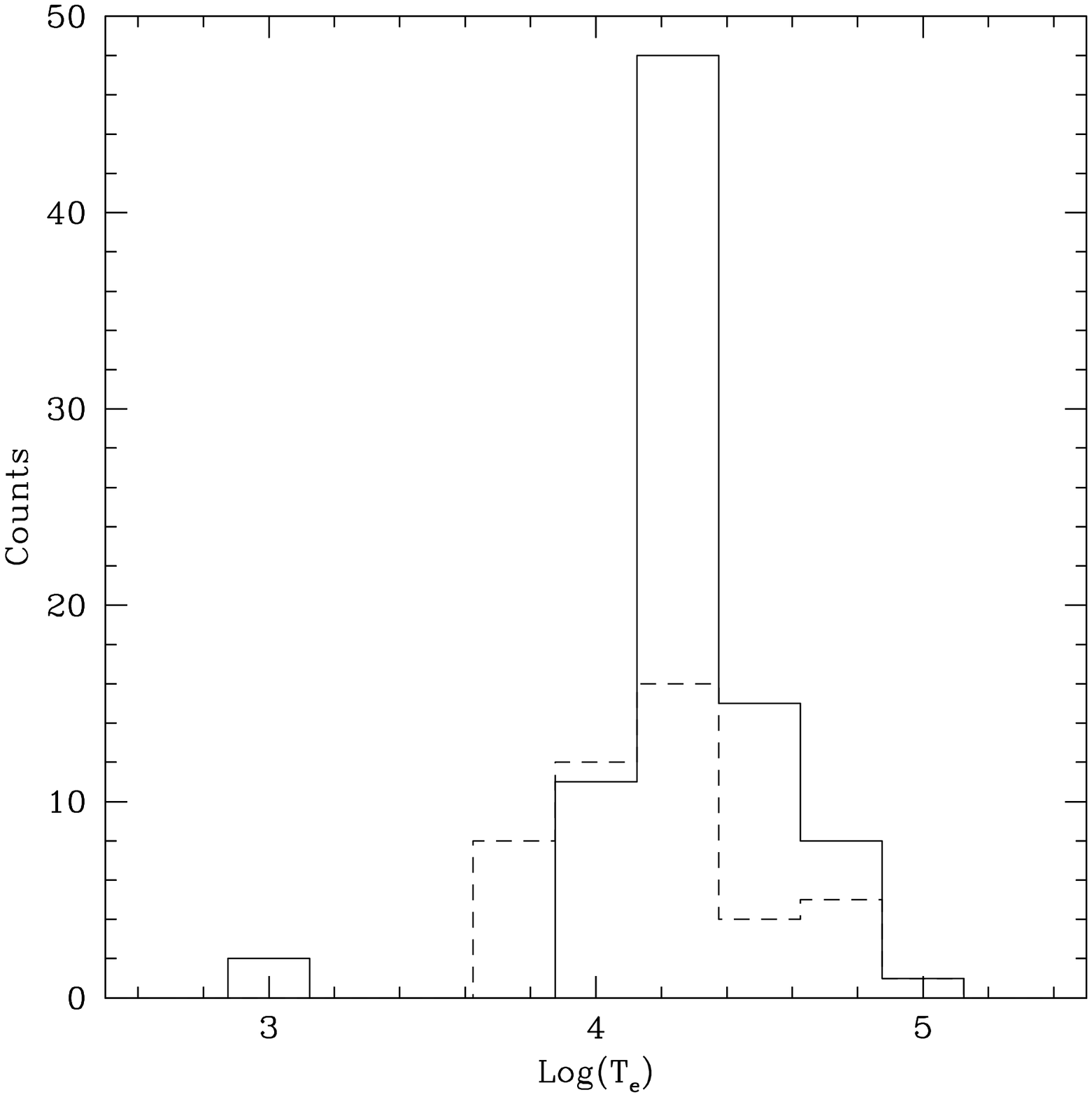}
\caption{Left: histogram of density obtained using a value of temperature $T\rm _e=10^4$ K, the median value is 450 cm$^{-3}$ (dashed line). Right: histograms of temperature obtained with [O\,{\sc iii}] (solid line) and [S\,{\sc ii}] (dashed line) ratio, the median values are $1.8 \times 10^4$ K and $1.5 \times 10^4$ K, respectively.}
\label{ist_t_den}
\end{figure}

We calculated the density with the {\sc temden} {\sc iraf} task, by using both the [S\,{\sc ii}] $\lambda$6716/$\lambda$6731 and the [Ar\,{\sc iv}] $\lambda$4711/$\lambda$4740 ratios. The [S\,{\sc ii}] and [Ar\,{\sc iv}] lines have different values of critical density ($\log N_{\rm e} =$ 3.2, 3.6, 4.4 and 5.6, respectively), thus we can assume that these lines are emitted by gas in different physical conditions. 
The [Ar\,{\sc iv}] ratio gives information about higher density and higher ionization gas, while lower density gas can be studied by means of the [S\,{\sc ii}] ratio.
We used an input value of temperature $T = 10^4$ K and we calculated the density for each spectrum in which we measured these ratios. Unfortunately, the [Ar\,{\sc iv}] lines could be measured for few central spectra only, giving not significant information about the spatial distribution of the high density gas. Therefore we calculated the density using the [S\,{\sc ii}] doublet, which is detectable in a more extended region, finding a median value of $\sim$ 450 cm$^{-3}$ (see Fig. \ref{ist_t_den}).
The central density obtained with [Ar\,{\sc iv}] is $\sim 2.2\times10^3$ cm$^{-3}$ (assuming $T\rm _e=10^4$ K) and is in good agreement with the value obtained with [S\,{\sc ii}] ($\sim 1.4\times10^3$ cm$^{-3}$). 
These higher values of density (1--1.4$\times 10^3 $ cm$^{-3}$) are in an internal region oriented roughly East-West.
\citet{2006A&A...456..953B} also estimated high density values ($>10^3$cm$^{-3}$) within an aperture of 2 arcsec on the nucleus (see their fig. 6, upper-right panel).
The gas temperature was obtained by calculating the ratio between the flux of the [O\,{\sc iii}] $\lambda\lambda$4959+5007 and the flux of the [O\,{\sc iii}] $\lambda$4363, 
and by calculating the ratio between the flux of [S\,{\sc ii}] $\lambda\lambda$6717+6731 and the flux of [S\,{\sc ii}] $\lambda\lambda$4068+4076. 
We determined the temperature with these two ratios, in case that the $\lambda$4363 and the $\lambda\lambda$4068+4076 emission lines could
 be measured. 
We derived the temperature using {\sc temden} and the values of density calculated with the [S\,{\sc ii}] ratio.
The histograms for the two determinations of temperature are shown in Fig. \ref{ist_t_den}. The median values are 1.8$\times 10^4$ K and 1.5$\times 10^4$ K for [O\,{\sc iii}] and [S\,{\sc ii}] ratios respectively. 
For high ionization gas, we should calculate the density using high ionization doublets, but in this case, we detected the [Ar\,{\sc iv}] emission lines in the inner parts only, and we could estimate the density using this ratio only for few nuclear regions, thus we used the [S\,{\sc ii}] density determination for the temperature estimate with the [O\,{\sc iii}] ratio. Higher values (up to 2--3 $\times10^4$ K) were found in the northern part of the FoV, while in the central regions we measured lower values (1--1.5 $\times10^4$ K) with [O\,{\sc iii}], and even less with [S\,{\sc ii}] (4--9 $\times10^3$K).

\begin{figure}
 \centering
\includegraphics[width=8.5cm]{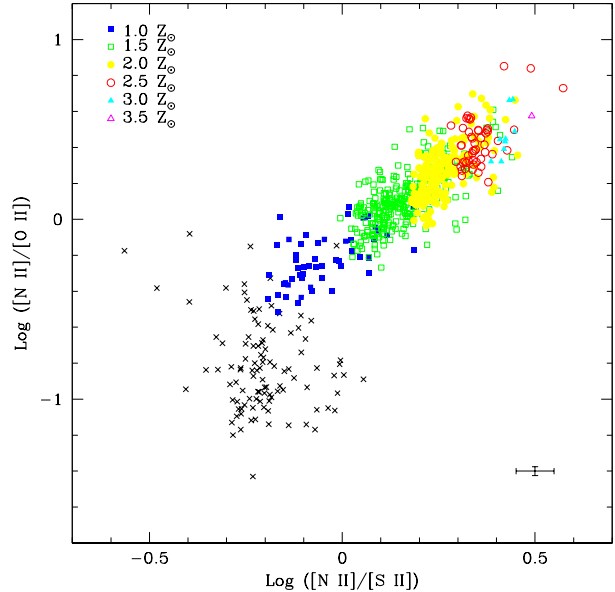}
\caption{The graph of the $\log$([N\,{\sc ii}]/[S\,{\sc ii}]) versus $\log$([N\,{\sc ii}]/[O\,{\sc ii}]) compared with theoretical models by \citet{Vaona2010} with D/G=1. The observed values are black crosses.}
\label{metallicita}
\end{figure}

In case of ionization by a power-law spectrum, as for an AGN, the determination of the metallicity is complicated because the ionization structure can be complex. There are not direct methods to determine the metallicity, and comparison with photoionization models must be used. 
To estimate the gas metallicity, we compare the measured fluxes with {\sc cloudy} models from \citet{Vaona2010}, using the emission line ratios [N\,{\sc ii}] $\lambda$6584/[O\,{\sc ii}] $\lambda$3727 vs. [N\,{\sc ii}] $\lambda$6584/[S\,{\sc ii}] $\lambda$6724.
Vaona showed that this is the more sensitive to metallicity diagram. The models have a low resolution in metallicity because he used few $\rm Z/Z_{\odot}$ values, nevertheless the trend is clear. 
We used models with dust-to-gas ratio (D/G) equal to 1.
The plot and the comparison between the observed values and the theoretical models allowed us to estimate a sub-solar metallicity for the ionized gas in NGC 7212 (see Fig. \ref{metallicita}). 
\begin{figure}
\includegraphics[width=4cm]{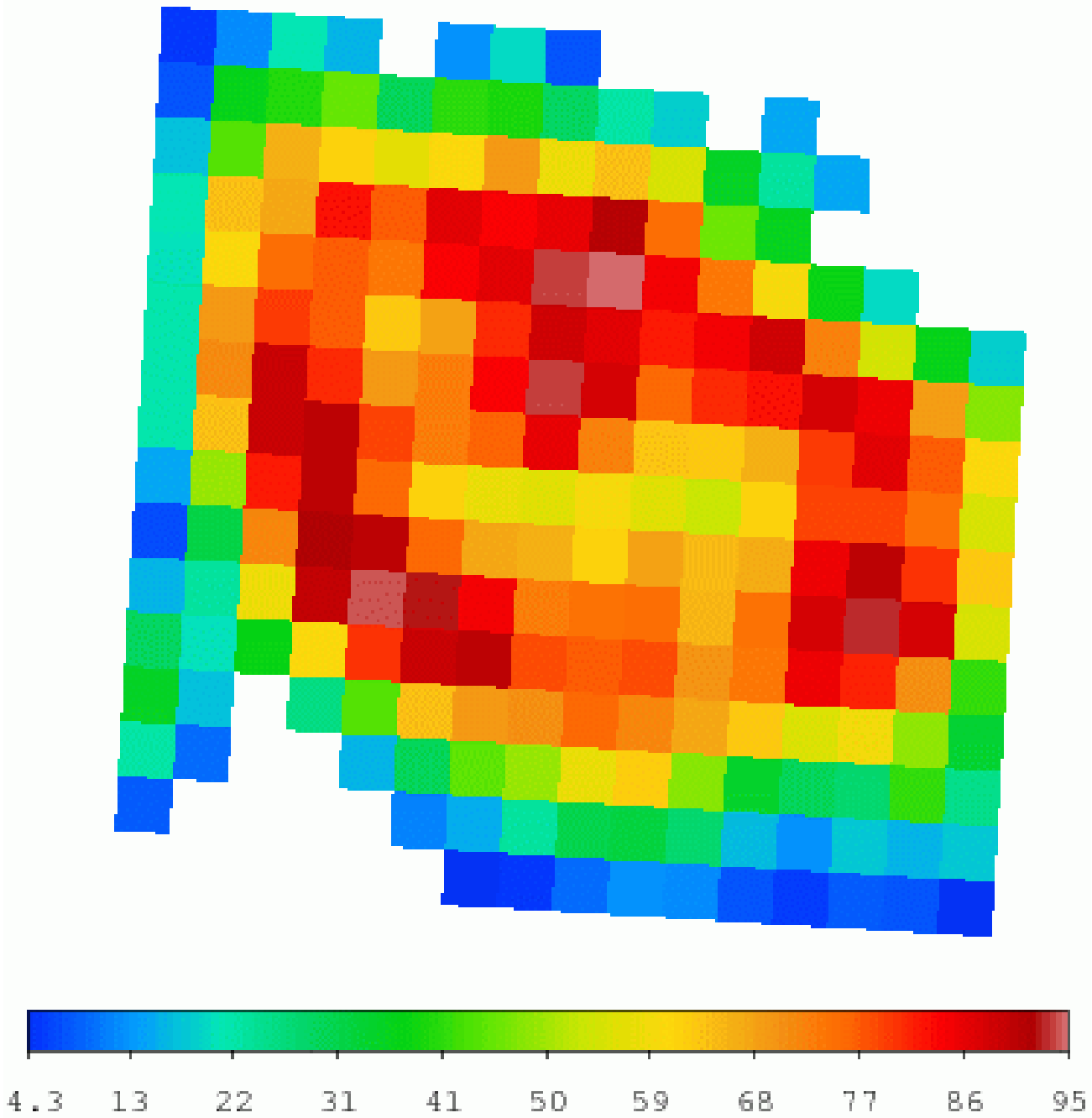}
\includegraphics[width=4cm]{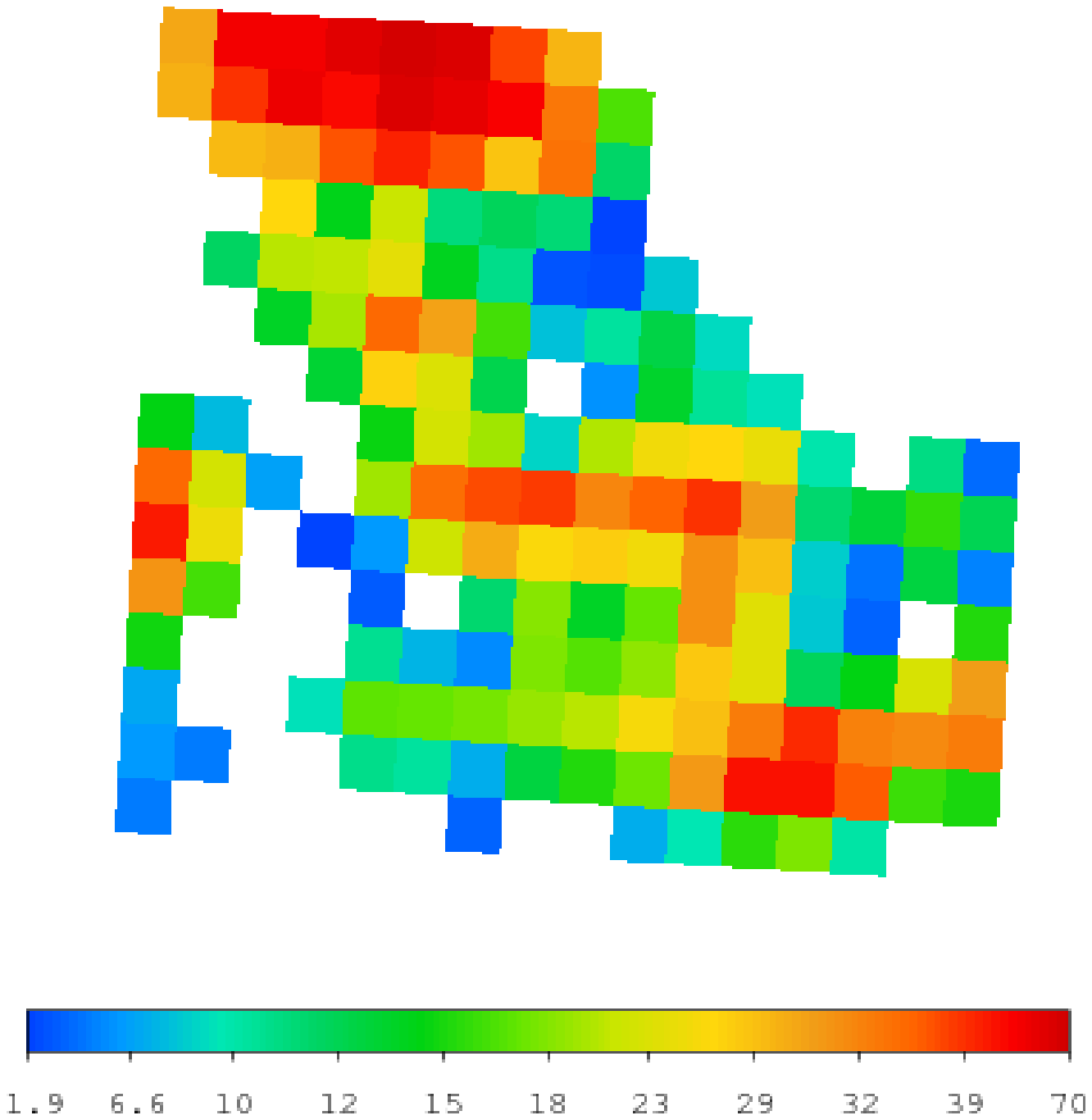}\\
\includegraphics[width=4cm]{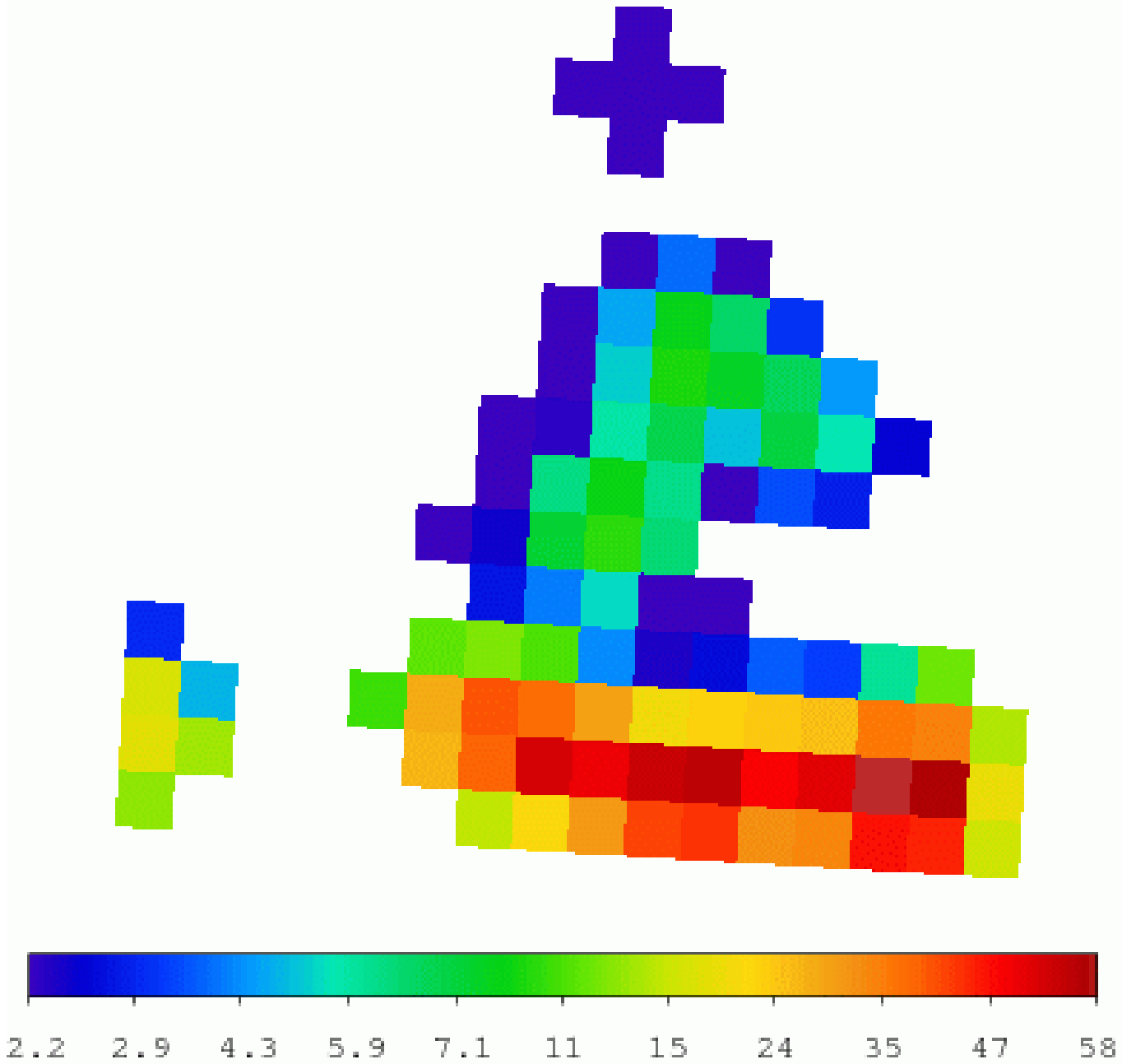}
\includegraphics[width=4cm]{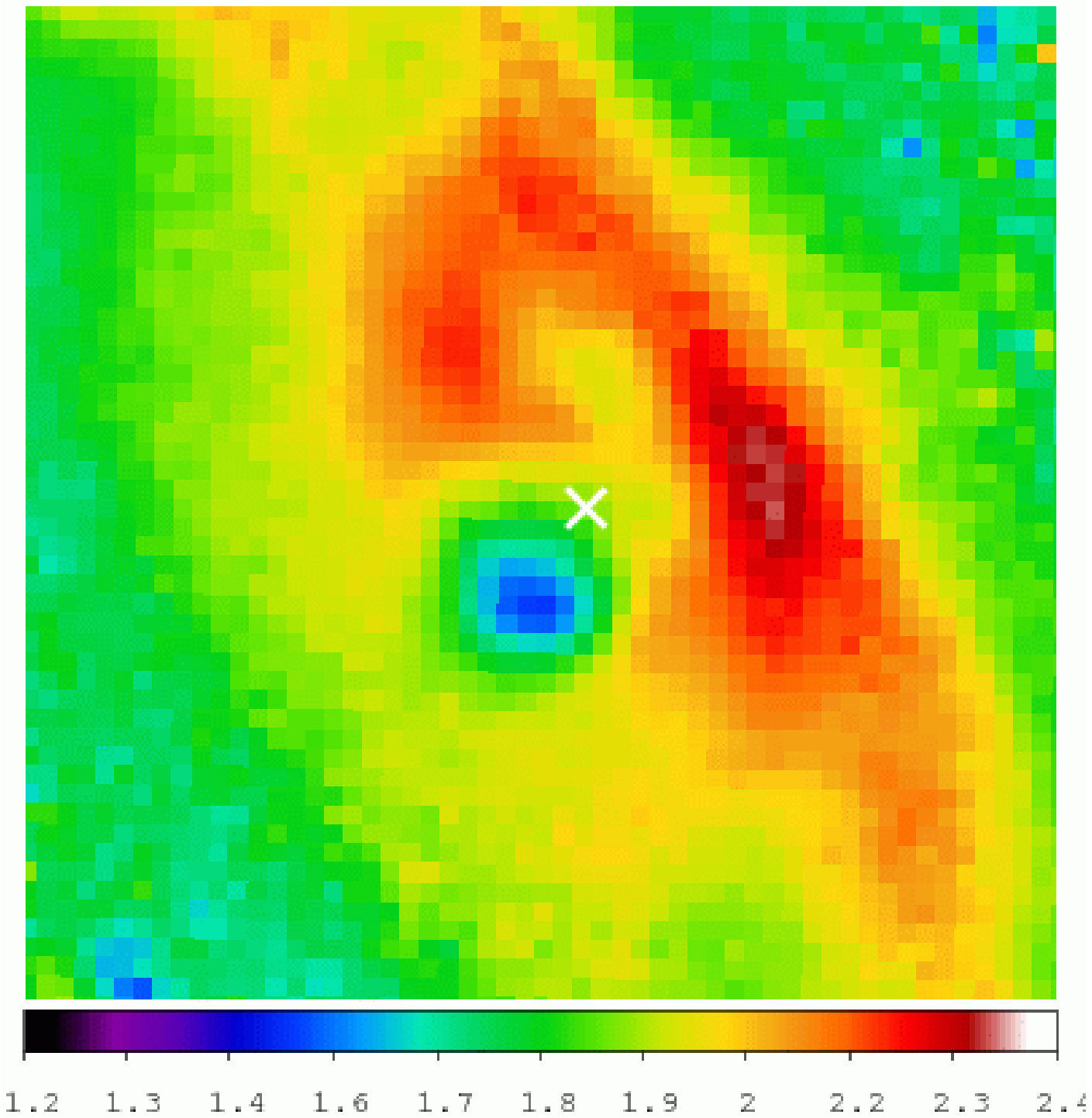}
\caption{Top: 2D maps of the old (left) and medium age (right) stellar population. Bottom: the 2D map of young population (left) and the $B-R$ color (right). The cross is the position of the galaxy nucleus. North is up and East is on the left.}
\label{starpop}
\end{figure}
\begin{figure*}
\centering
\includegraphics[width=7.4cm]{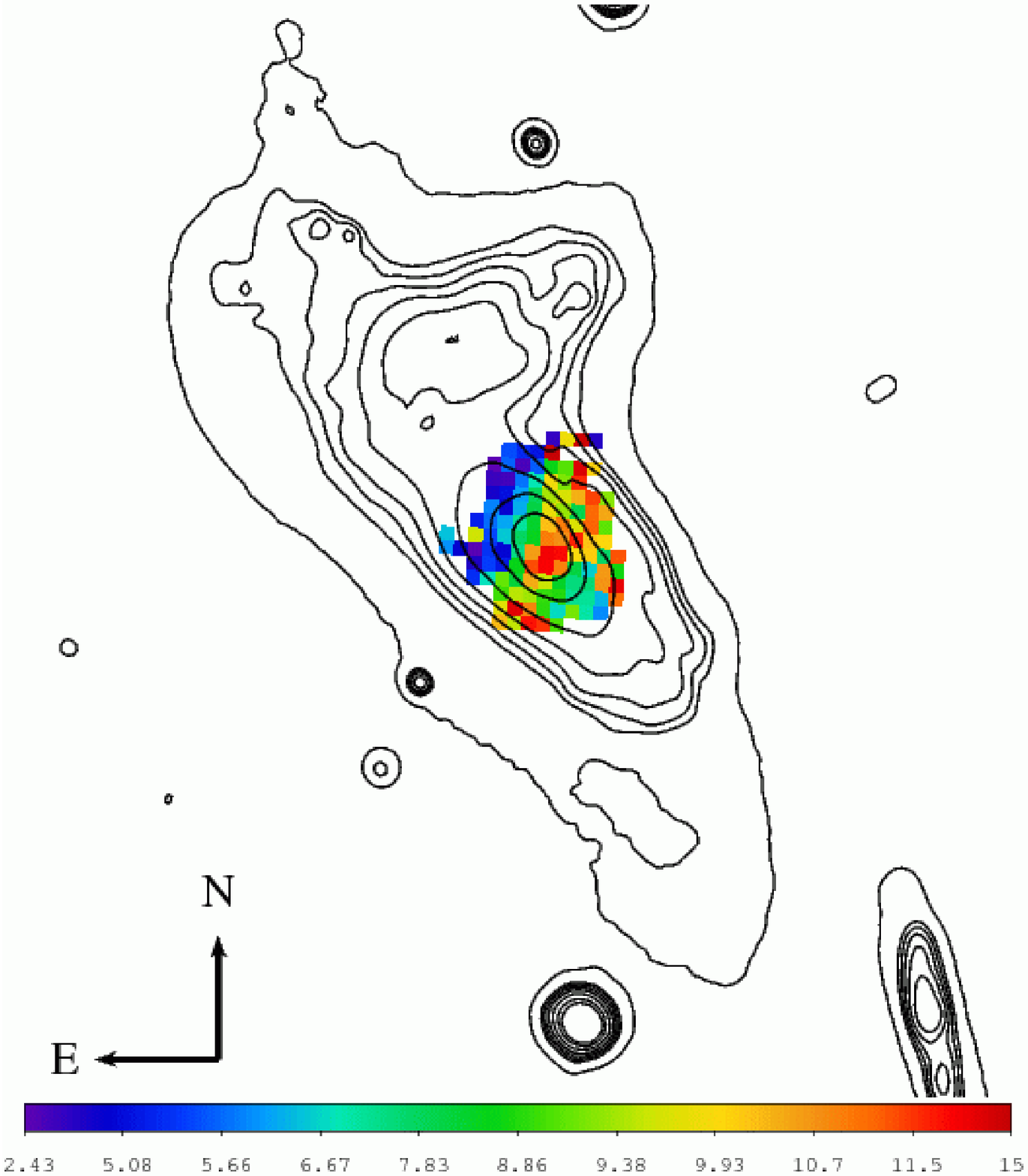}
\includegraphics[width=7.4cm]{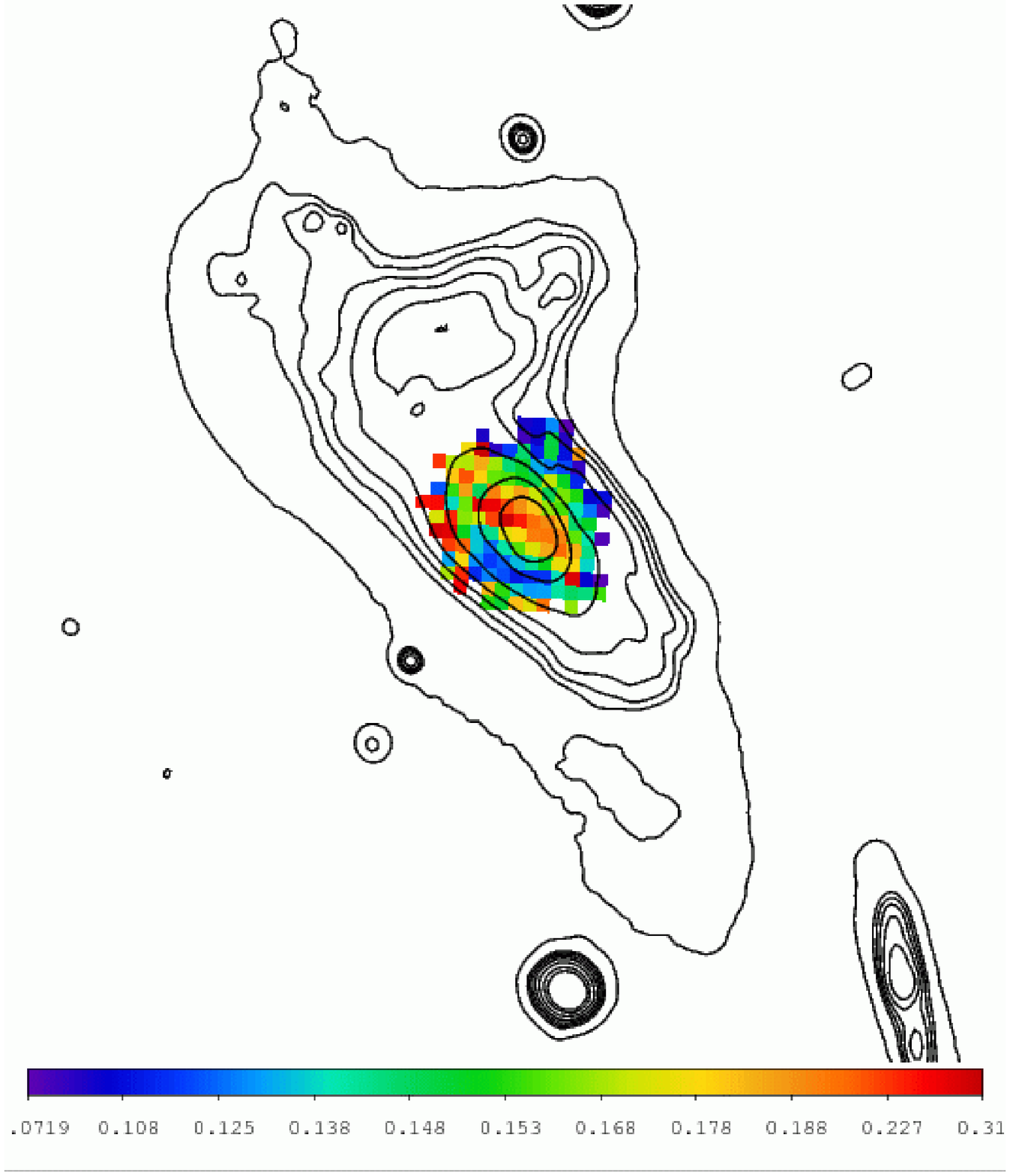}\\
\includegraphics[width=7.4cm]{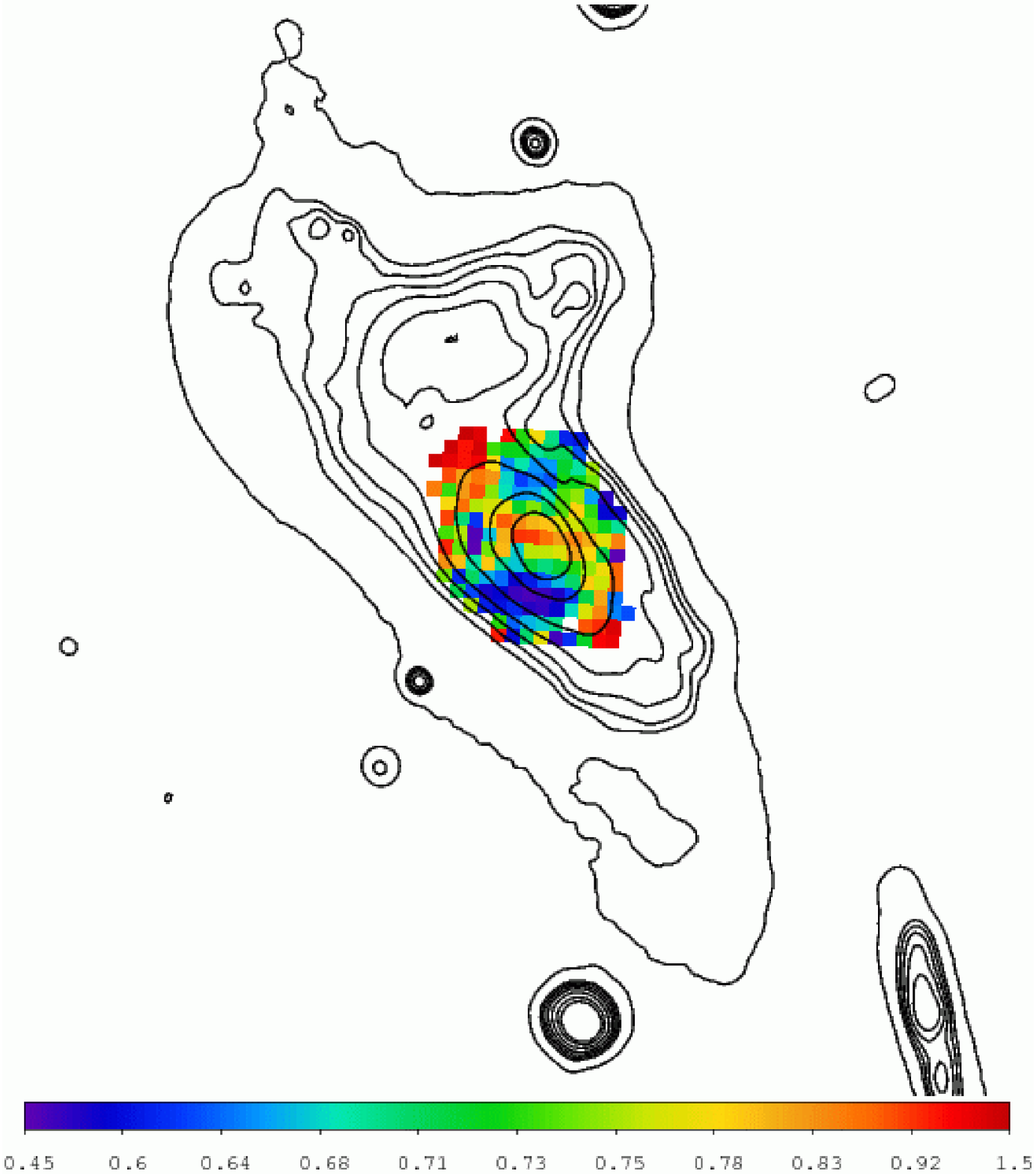}
\includegraphics[width=7.4cm]{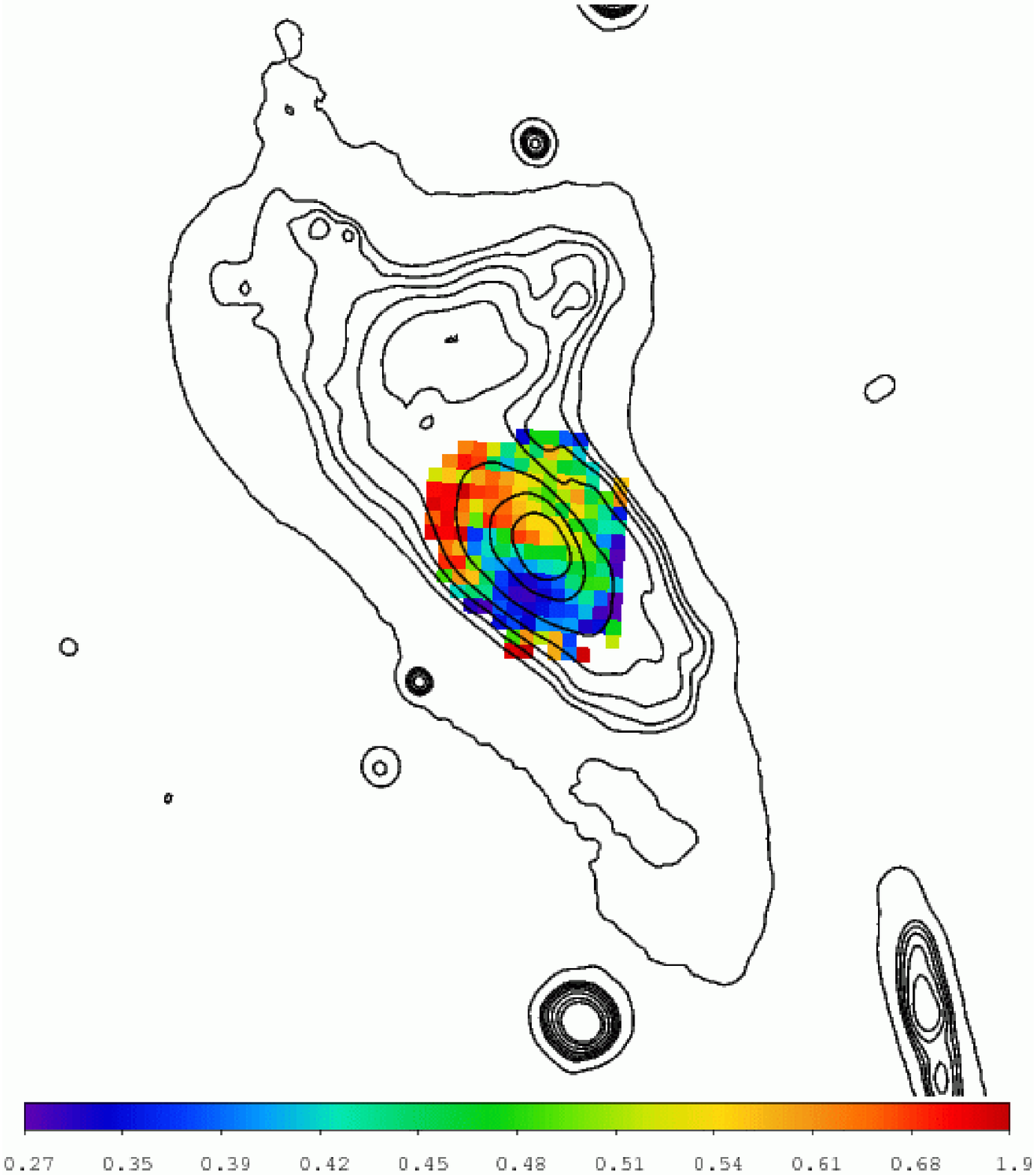}
\caption{Maps of diagnostic diagrams with the $R$-band SAO contours overlaid (in black). Top: [O\,{\sc iii}]/H$\beta$ (left) and [O\,{\sc i}]/H$\alpha$ (right). Bottom: [N\,{\sc ii}]/H$\alpha$ (left) and [S\,{\sc ii}]/H$\alpha$ (right).}
\label{diagnostic_maps}
\end{figure*}

From the {\sc starlight} output, we reconstructed the 2D maps of the stellar velocity field, as well as the 2D maps of the distribution of the stellar population with different ages and metallicities. 
The maps were built using only the models in which the \textit{adev} parameter was less than 50 per cent. This parameter is an output of {\sc starlight} and corresponds to the average percentage value ($|O_{\lambda}-M_{\lambda}|/O_{\lambda}$) of the deviation between the observed spectrum ($O_{\lambda}$) and the modelled one ($M_{\lambda}$) over all fitted pixels. 
Then we selected only the spectra with $\rm S/N>3$. 
We obtained the flux distributions for stellar populations in different ranges of age: young ($t < 10^8$ yr), medium age (10$^8$yr $< t <$ 10$^{9.5}$yr), and old ($t > 10^{9.5}$yr); and with different metallicities: sub-solar (Z = 0.004), solar (Z = 0.02) and super-solar (Z = 0.05).
We created the $B-R$ colour image from the broad-band images obtained at SAO and we compared it with the 2D maps showing the stellar population distribution. 
The colour image (Fig. \ref{starpop}, bottom-right panel) shows a red ring-like structure North and North-West of the nucleus, corresponding to the large dust lane already visible in the broad-band images and in our A($V$) map, and a blue region South-East of the nucleus. 
Both these structures were detected by \citet{1998A&AS..132..197K}, who found a blue fan-shaped region along PA = 165\degr\ extended about 2.3 arcsec. Our region has the same size but its shape is circular due to our seeing of about 1.5 arcsec, worst than the 0.7-arcsec seeing in the Kotilainen images.
As in \citet{2003MNRAS.339..772R} we found that the old population ($t> 10^{9.5}$yr) is the main component and in the nucleus we found also a younger component ($10^8$yr $ < t < 10^{9.5}$yr).
In particular, in a large ring-like structure the 70--80 per cent of the total light is due to the old stellar population. A lower precentage of the light (50--60 per cent) is observed in a central elongated region oriented at PA = 90\degr, where the intermediate age population accounts for 30--40 per cent of the total. This apparent hollow in the distribution of the light of the old stars is simply an effect of the presence of younger stars with smaller $M/L$ ratio. Finally, about 10 per cent of the light seems to be emitted by young population, but we stress that it could be an effect of diffuse light from the cone and scattered light from the AGN. In fact {\sc starlight} does not distinguish between OB stars contribution and a power-law continuum. Therefore, even if in a Seyfert 2 galaxy the AGN continuum is strongly absorbed with respect to Seyfert 1, the detection of the young and very young population could be wrong, due to the contamination from the AGN featureless continuum.In addition, high values of intermediate stellar population are found in the interaction regions (North-East of the nucleus). This component could result from the on-going merger.

We used the diagnostic ratios, applied in the BPT and VO diagrams \citep{1981PASP...93....5B,1987ApJS...63..295V}, to investigate the ionization mechanisms in act in the ENLR.
The 2D maps of the diagnostic ratios are shown in Fig. \ref{diagnostic_maps}. The [N\,{\sc ii}]/H$\alpha$, [O\,{\sc i}]/H$\alpha$ and [S\,{\sc ii}]/H$\alpha$ maps were created using the non-reddening corrected values, in order to study a more extended field, not being limited by the small size of the H$\beta$ emission.
From the 2D maps, we detected for the first time an extended structure with high values of [O\,{\sc iii}]/H$\beta$, larger than 10, suggesting the possible presence of an ionization cone oriented as the galaxy minor axis (see Section \ref{ioniz_cones}). 
The [O\,{\sc iii}]/H$\beta$ ionization map shows high values even far from the nucleus.
If photoionization is the only ionization mechanism, the photon flux is diluted with increasing distance, and it is not expected to remain high at large distance from the nuclear source. 
We speculate that this could be a density effect, in fact the ionization parameter $\rm U=\frac{Q_{ion}}{cr^2N_H} $, where $\rm Q_{ion}$ is the number of ionizing photons per second emitted by the source, depends both on the inverse of the density and of the distance. 
So high values of U could be due to high values of ionizing photon flux or to low density. Inside the cone, we found low density, but without a gradient, the value is approximately the same along the whole size of the cone. Thus, we can exclude that high values of U are due to low density. 
We can invoke the presence of shocks as an additional mechanism of ionization. Indeed, some regions inside the cone show electron temperature values larger than $3\times10^4$ K, suggesting that shocks could be at work. However, this hypothesis will be tested with photoionization+shocks models (Contini et al. 2011, in preparation).

\begin{figure}
\centering
 \includegraphics[width=9cm]{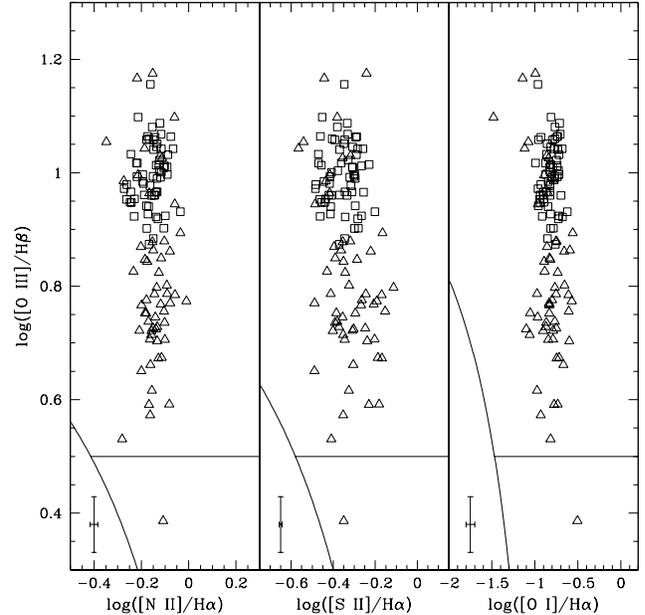}
\caption{The VO diagnostic diagrams. The squares are regions inside the cone, the triangles are regions outside.}
\label{diagnostici_tot}
\end{figure}

In addition, according to the very high values of [O\,{\sc iii}]/H$\beta$, we can also exclude a contribution from star formation. 
The diagnostic ratios of [N\,{\sc ii}]/H$\alpha$\ and [S\,{\sc ii}]/H$\alpha$\ are not smoothly distributed, showing higher values (respectively $\sim$1--1.3 and $\sim$0.7--0.9) orthogonal to the cone and outside it. Furthermore, these ratios show higher values in the North-Eastern regions, namely in the interacting regions between the two galaxies, as clearly seen in Fig. \ref{diagnostic_maps} with the galaxy contours overlaid. 
Here the ratios can be strongly influenced by interaction effects, such as shocks. 
The [O\,{\sc i}]/H$\alpha$\ map is peculiar. It shows an elongated shape along the galaxy major axis, with high values oriented as the high density structure. 
High values of [O\,{\sc i}]/H$\alpha$ are related to collisions.
The [O\,{\sc i}] emission line is formed in the recombination region, where O$^+$, H$^+$ and H exist. This region is more extended if the ionizing source has a power-law spectrum. This ratio is very useful to distinguish between photoionization by AGN or stars. 
We plotted the VO diagnostic diagrams, using the reddening corrected fluxes, both for the regions inside and outside the ionization cones (see Fig. \ref{diagnostici_tot}). 
In order to separate the different regions in the diagnostic diagram (AGN, H\,{\sc ii} regions and LINERs), we used the expressions from \citet{2006MNRAS.372..961K}. By analysing these diagrams we noticed that 
all the regions occupy the AGN area and are distributed vertically towards the shock region. This means that for each region, inside and outside the cone, the ionizing source has a power-law spectrum. The only difference is that for the regions outside the cone the ionization decreases with increasing radius, instead, inside the cone, it remains high also far from the nucleus (see Fig. \ref{O3Hb_dist}). Possible explanations are that the ionizing photons from the cone are diffused, or that the cone aperture angle is larger.

\section[]{Kinematics}
\begin{figure}
\centering
\includegraphics[width=4cm]{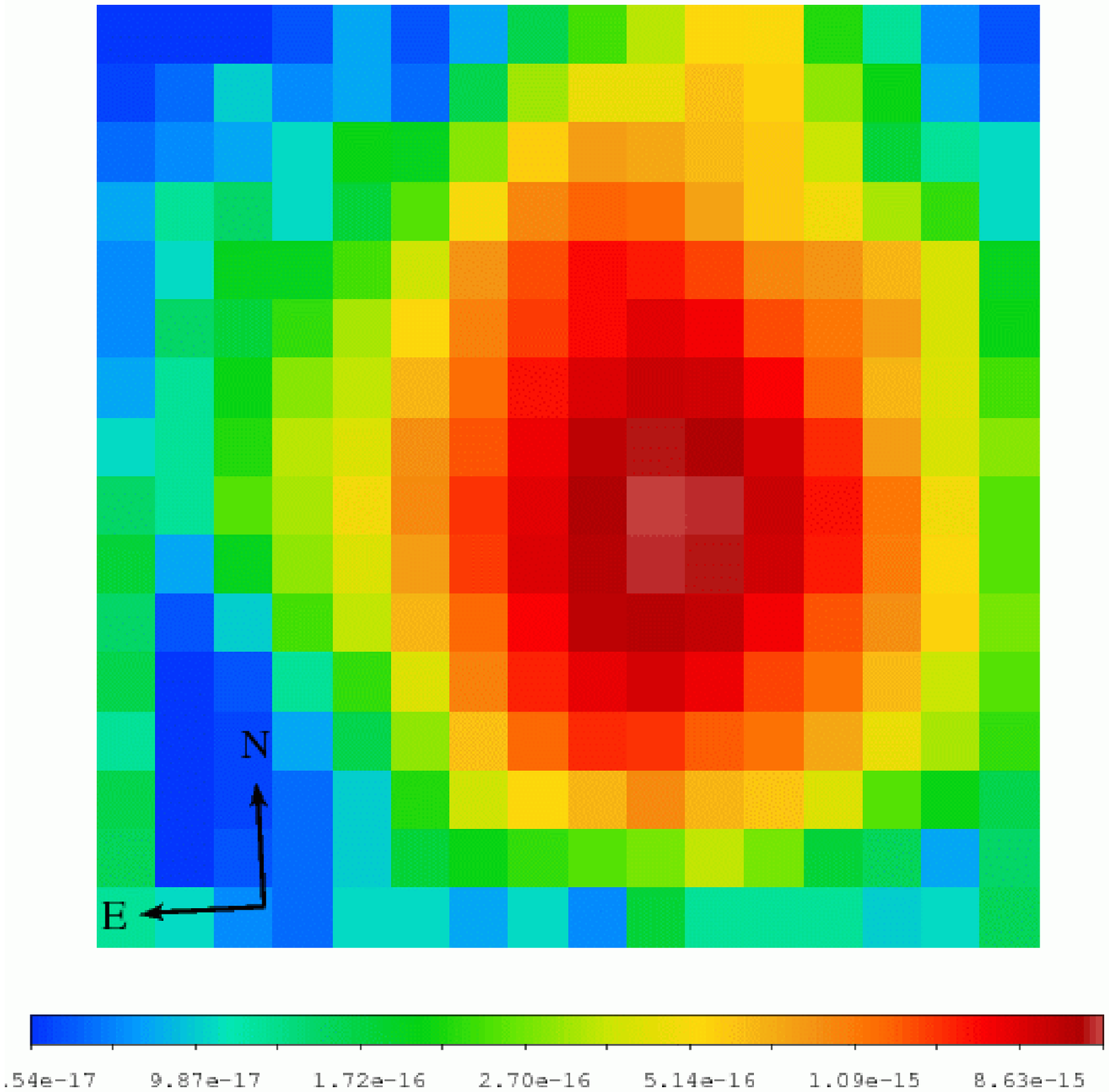}
\includegraphics[width=4cm]{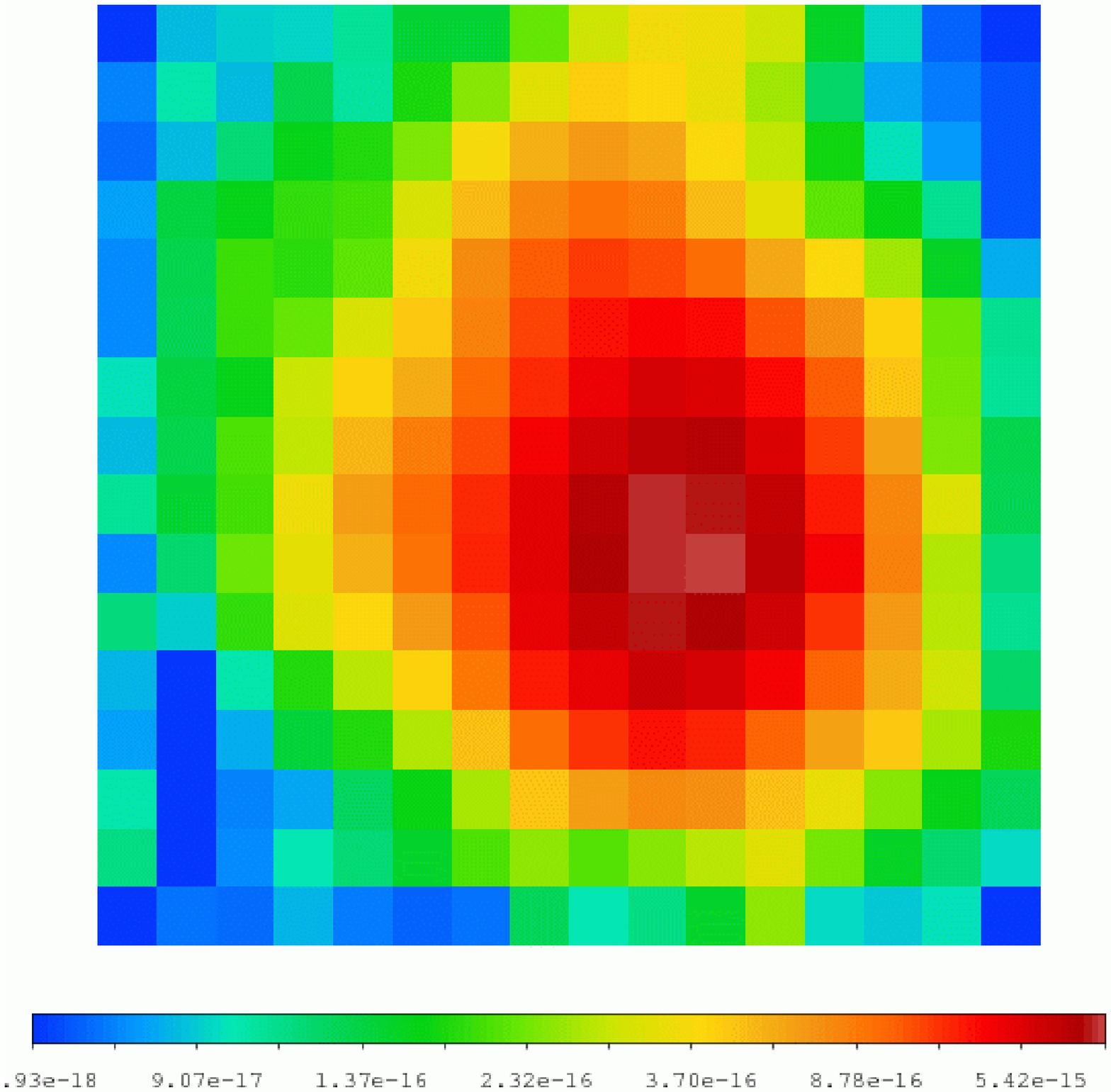}\\
\includegraphics[width=4cm]{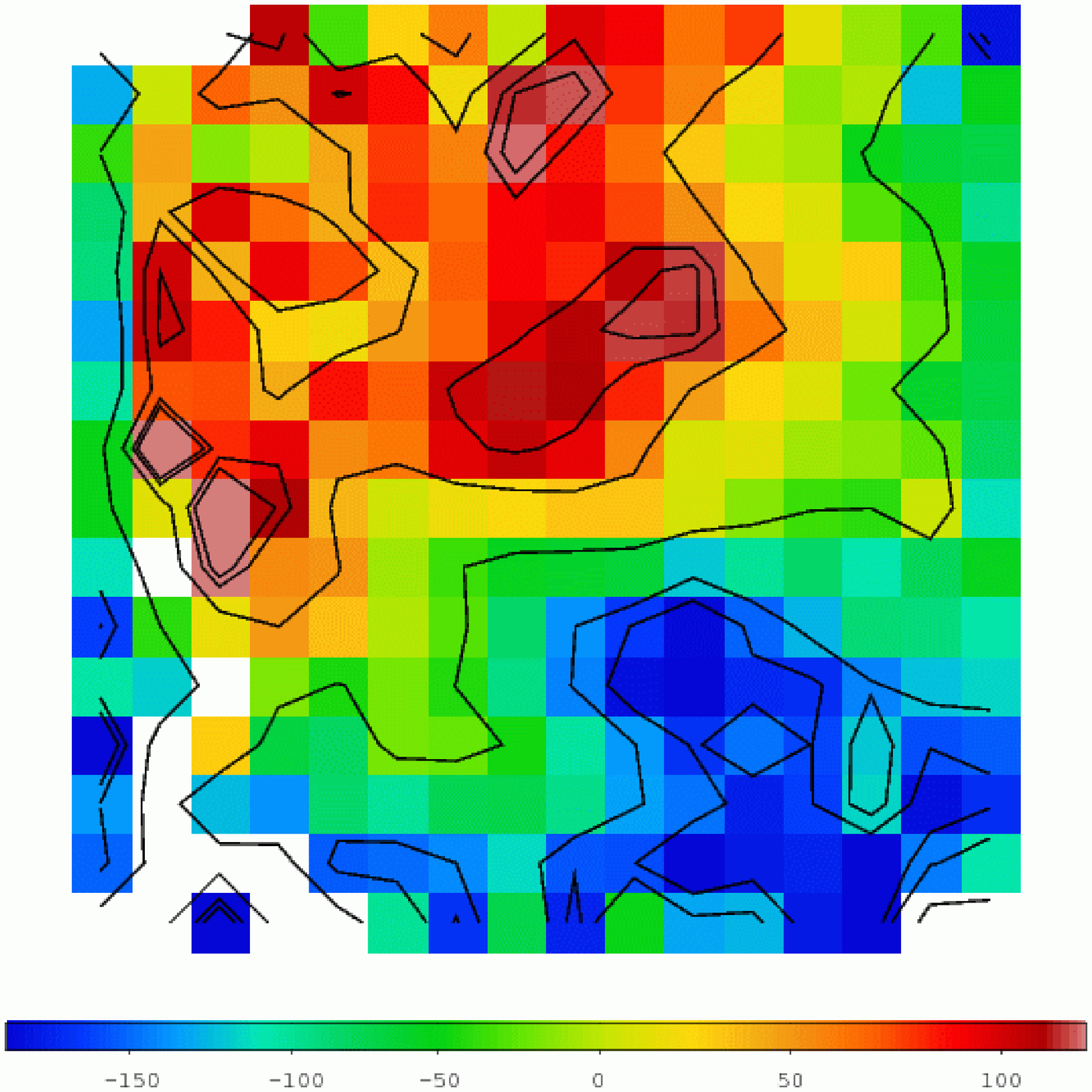}
\includegraphics[width=4cm]{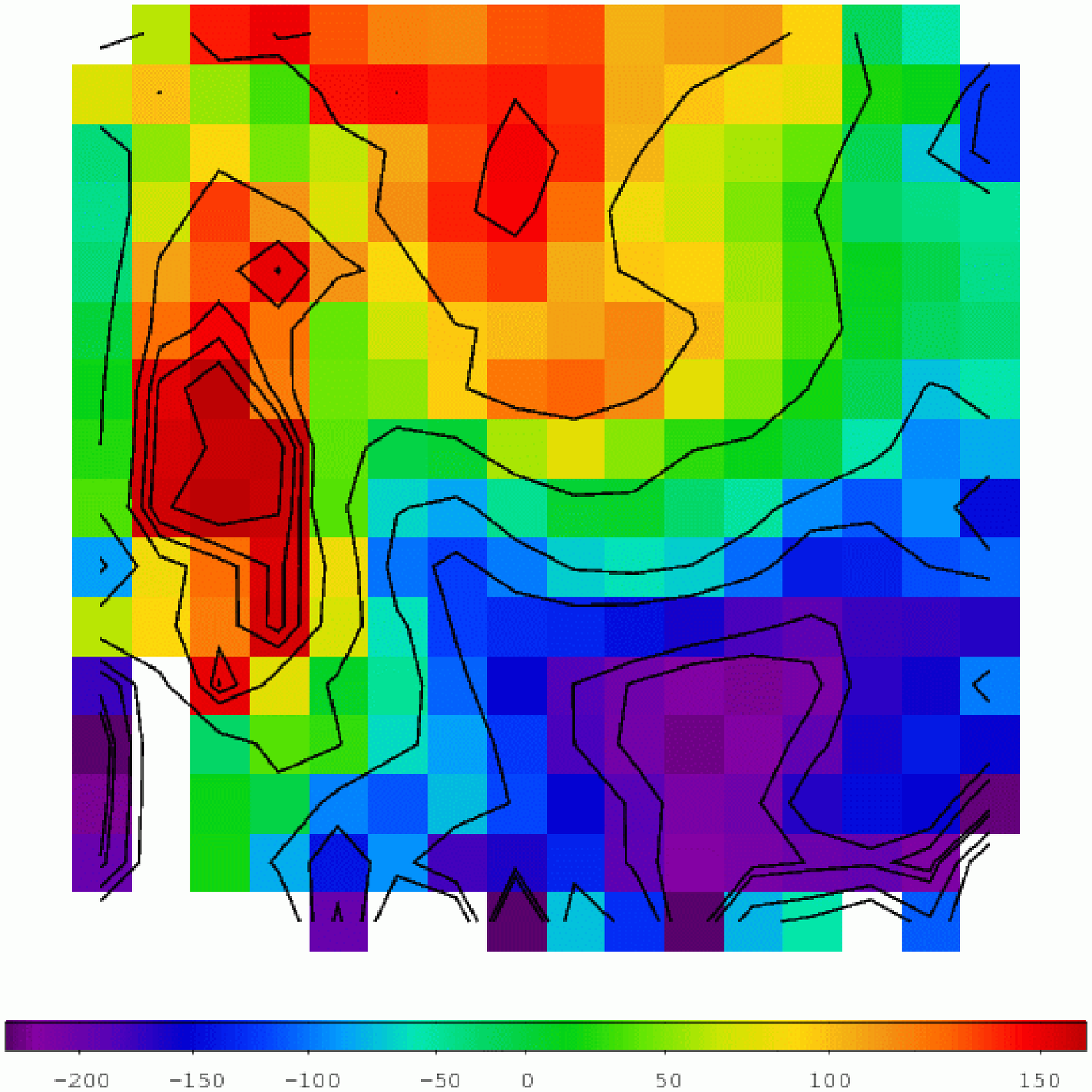}\\
\includegraphics[width=4cm]{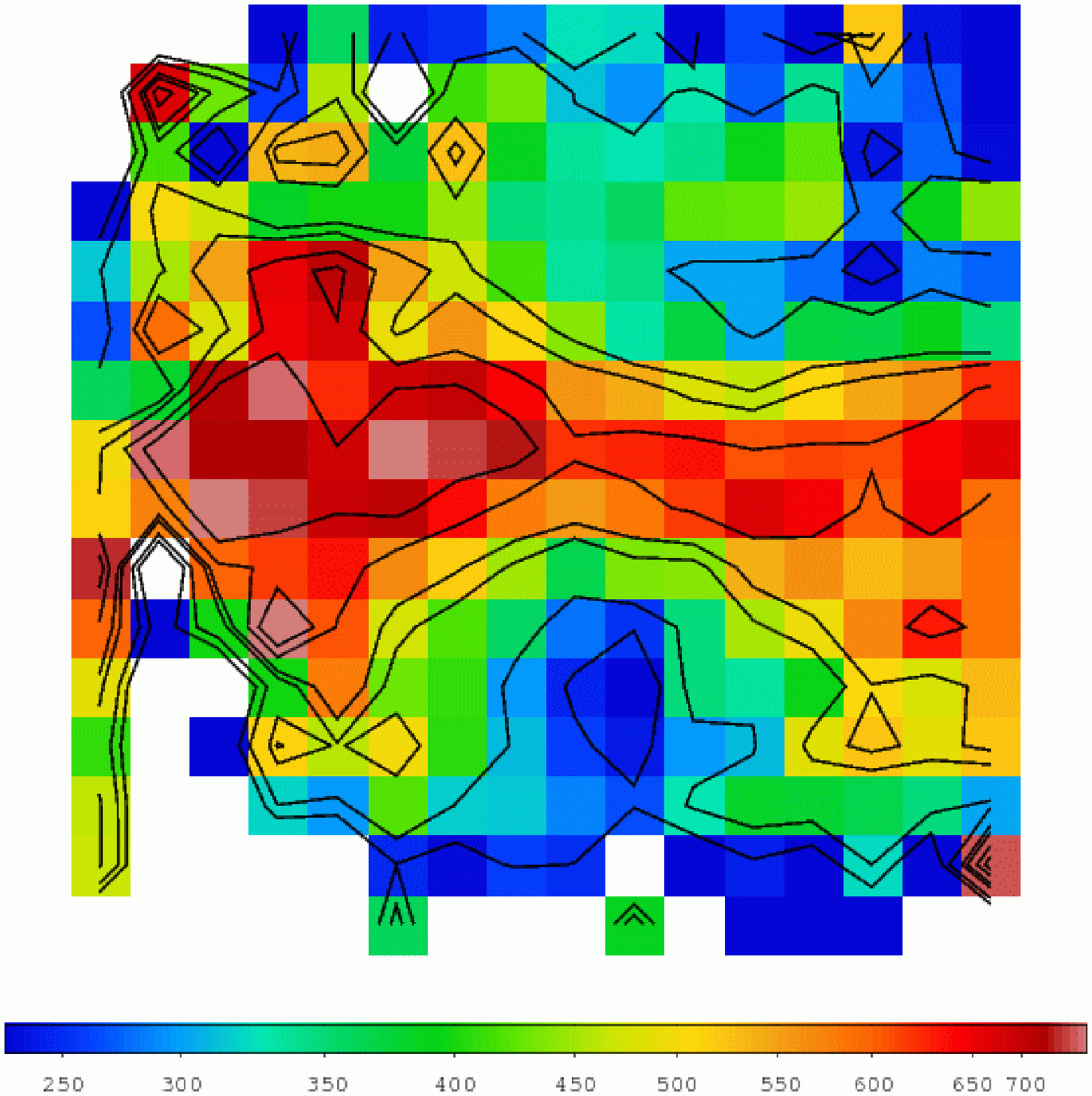}
\includegraphics[width=4cm]{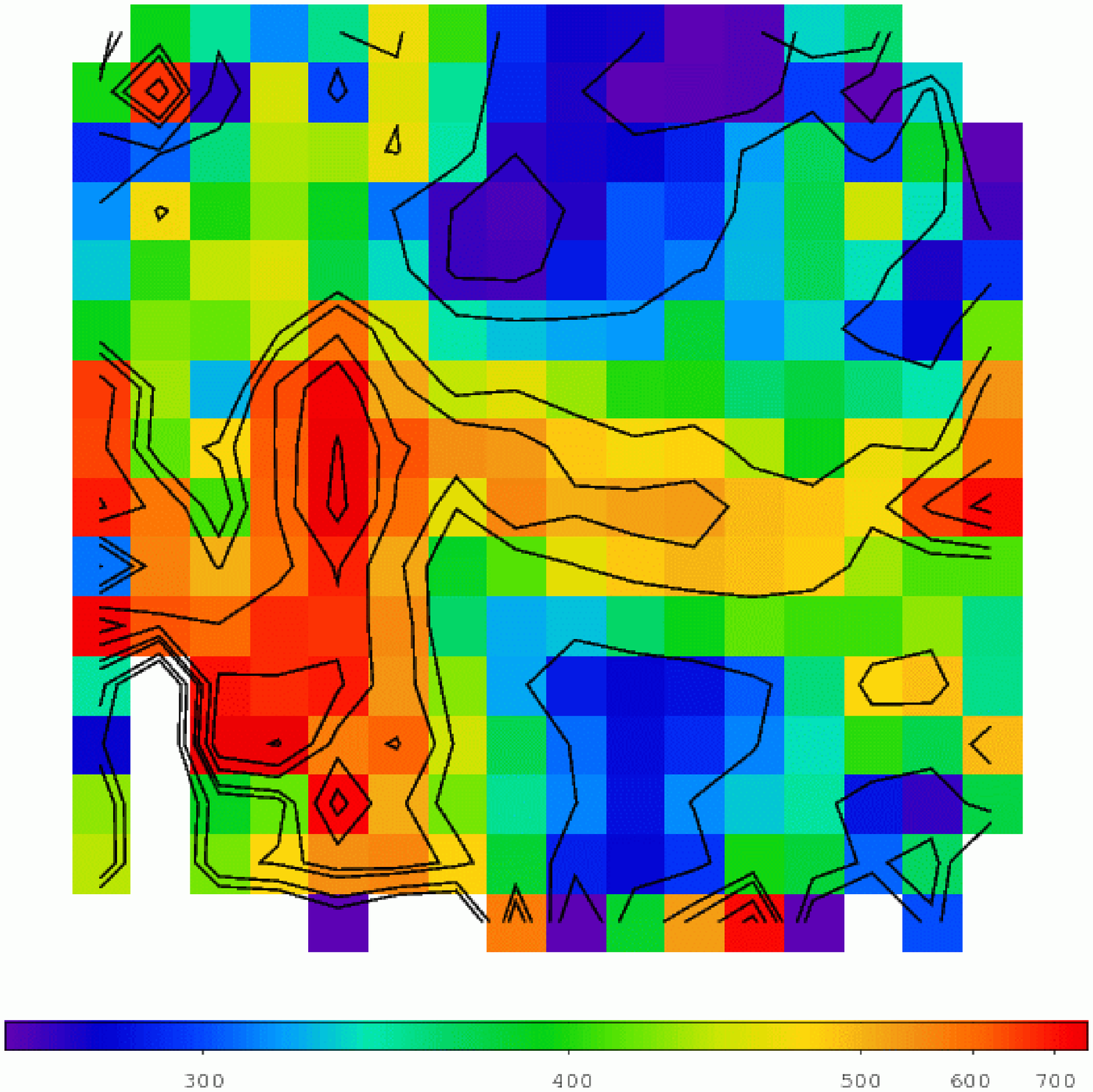}
\caption{Left: 2D maps of flux (top), velocity (middle) and FWHM (bottom) for high ionization gas ([O\,{\sc iii}]). Right: the same for low ionization gas (H$\alpha$).}
\label{v600}
\end{figure}

\begin{figure}
 \centering
\includegraphics[width=2.7cm]{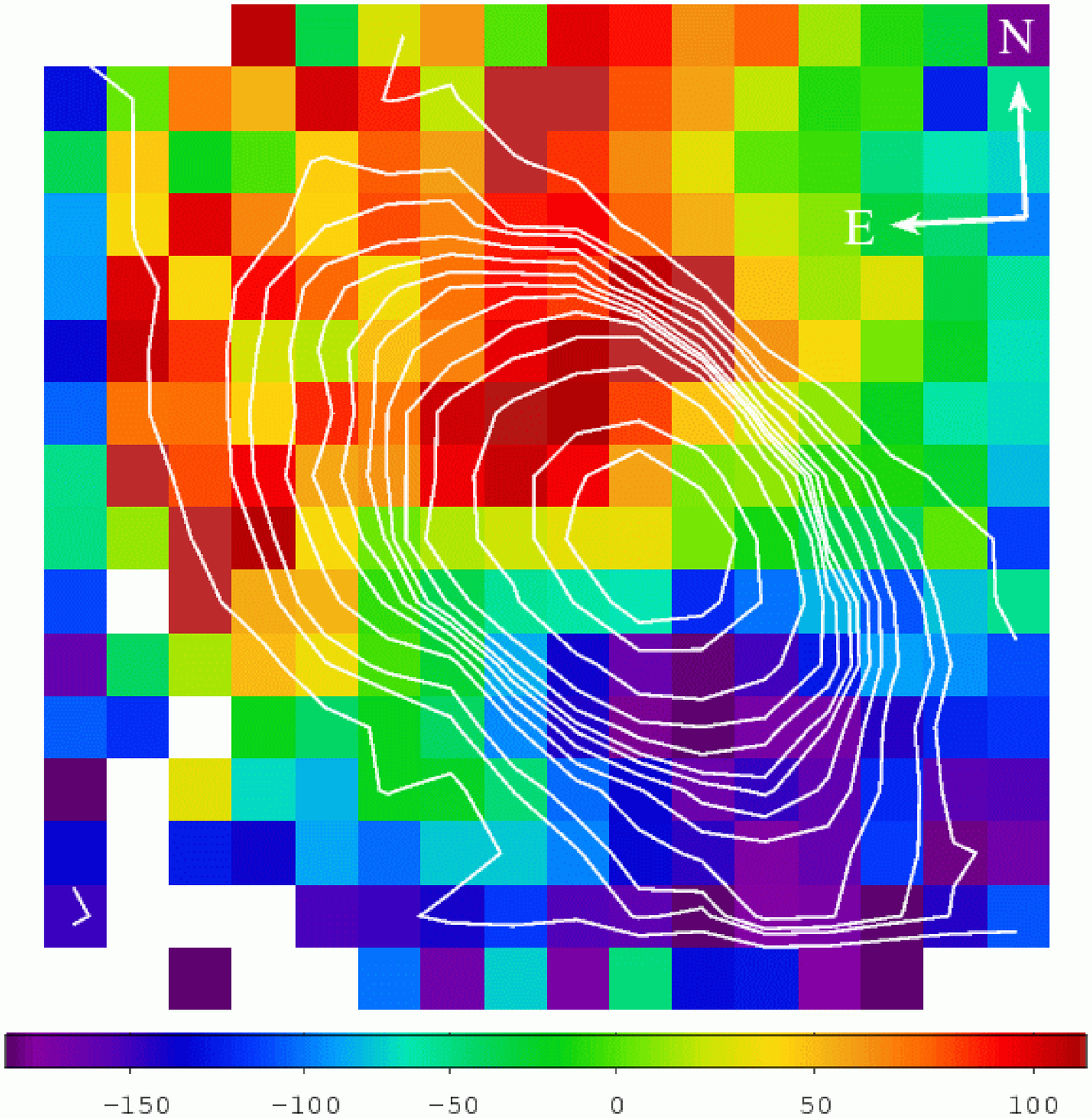}
\includegraphics[width=2.7cm]{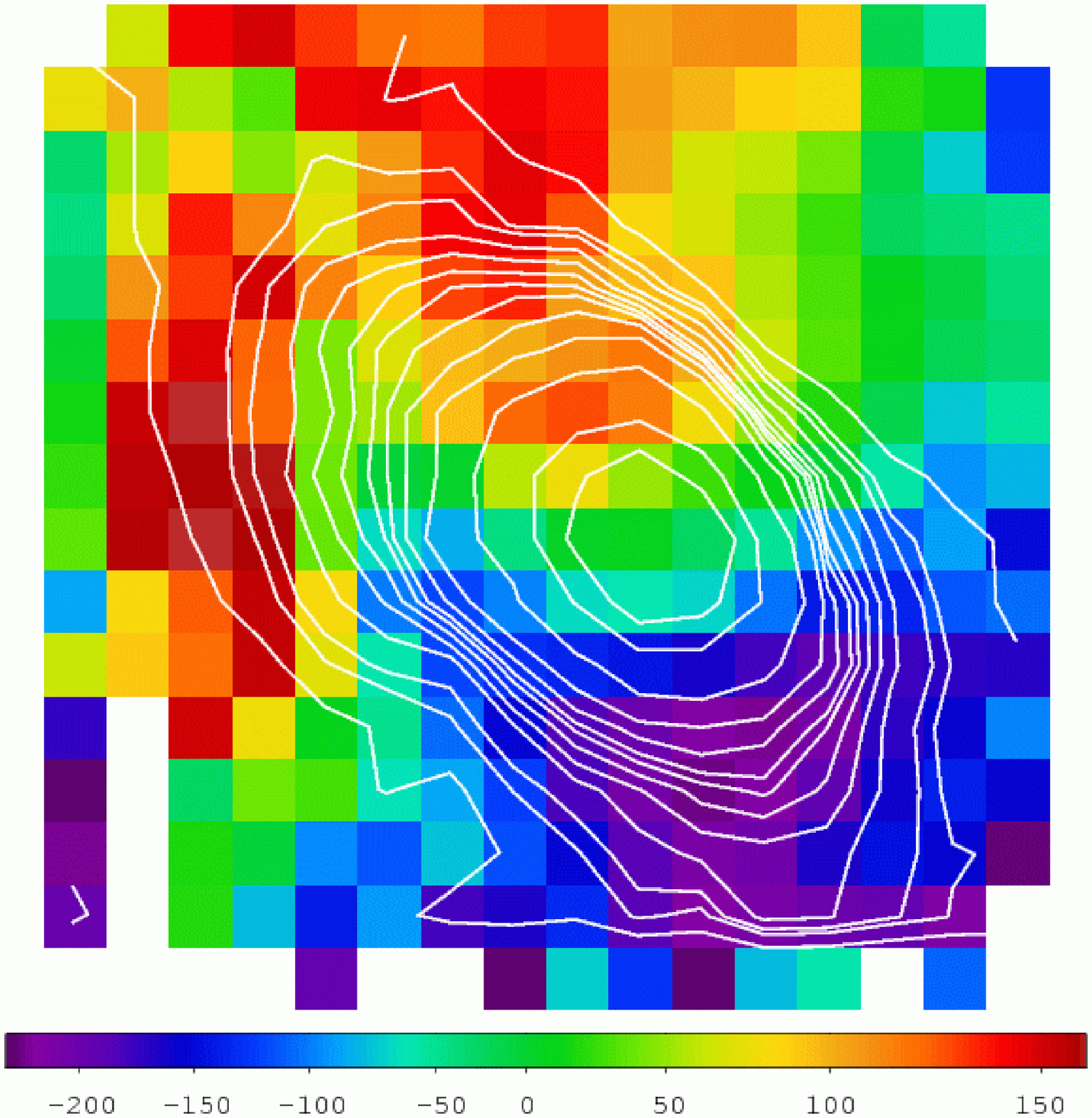}
\includegraphics[width=2.7cm]{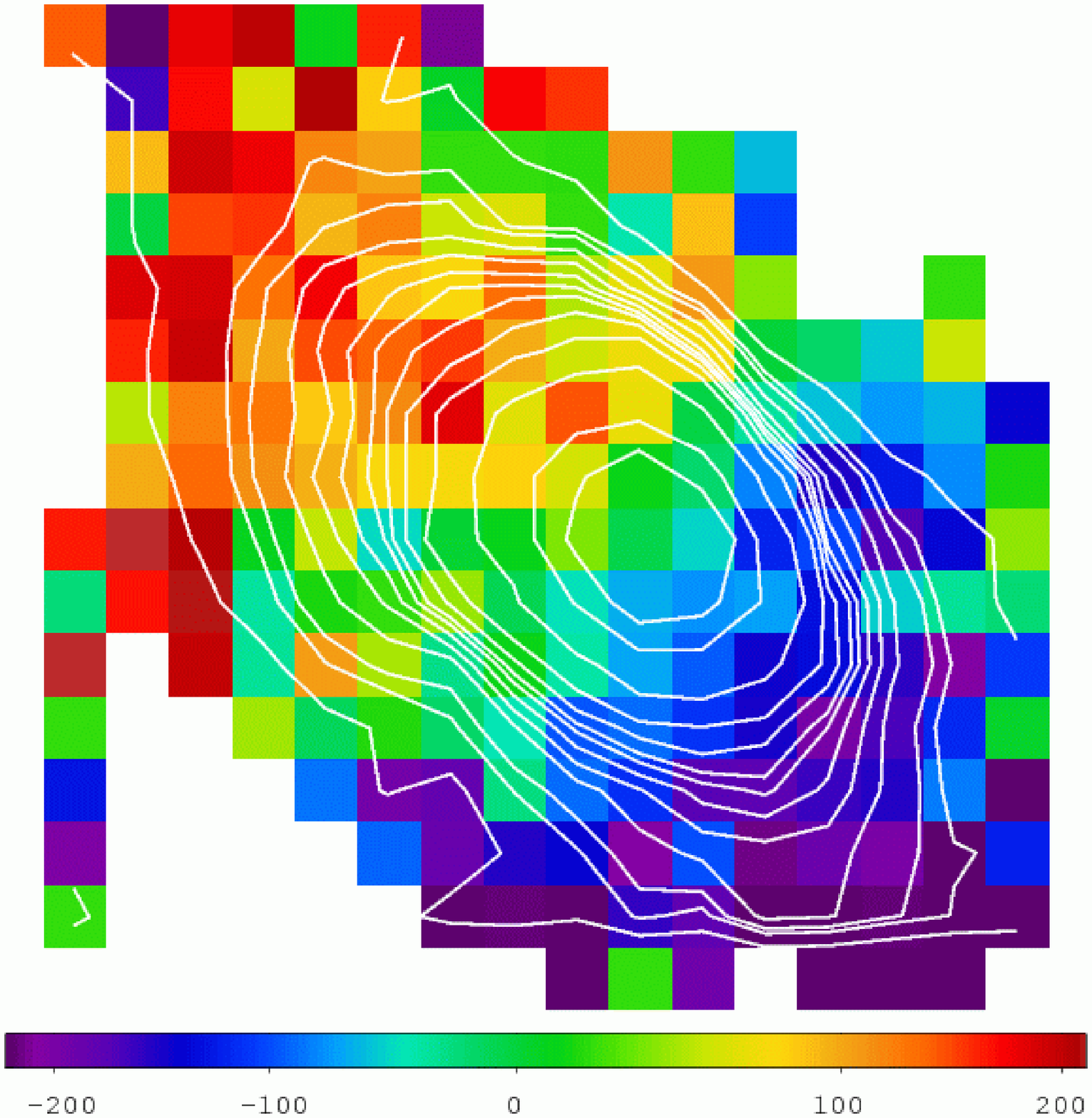}
\caption{The velocity maps of [O\,{\sc iii}] (left), H$\alpha$ (middle) and stars (right) with the continuum contours overlaid.}
\label{v_gas_star}
\end{figure}

\begin{figure}
 \centering
\includegraphics[width=4cm]{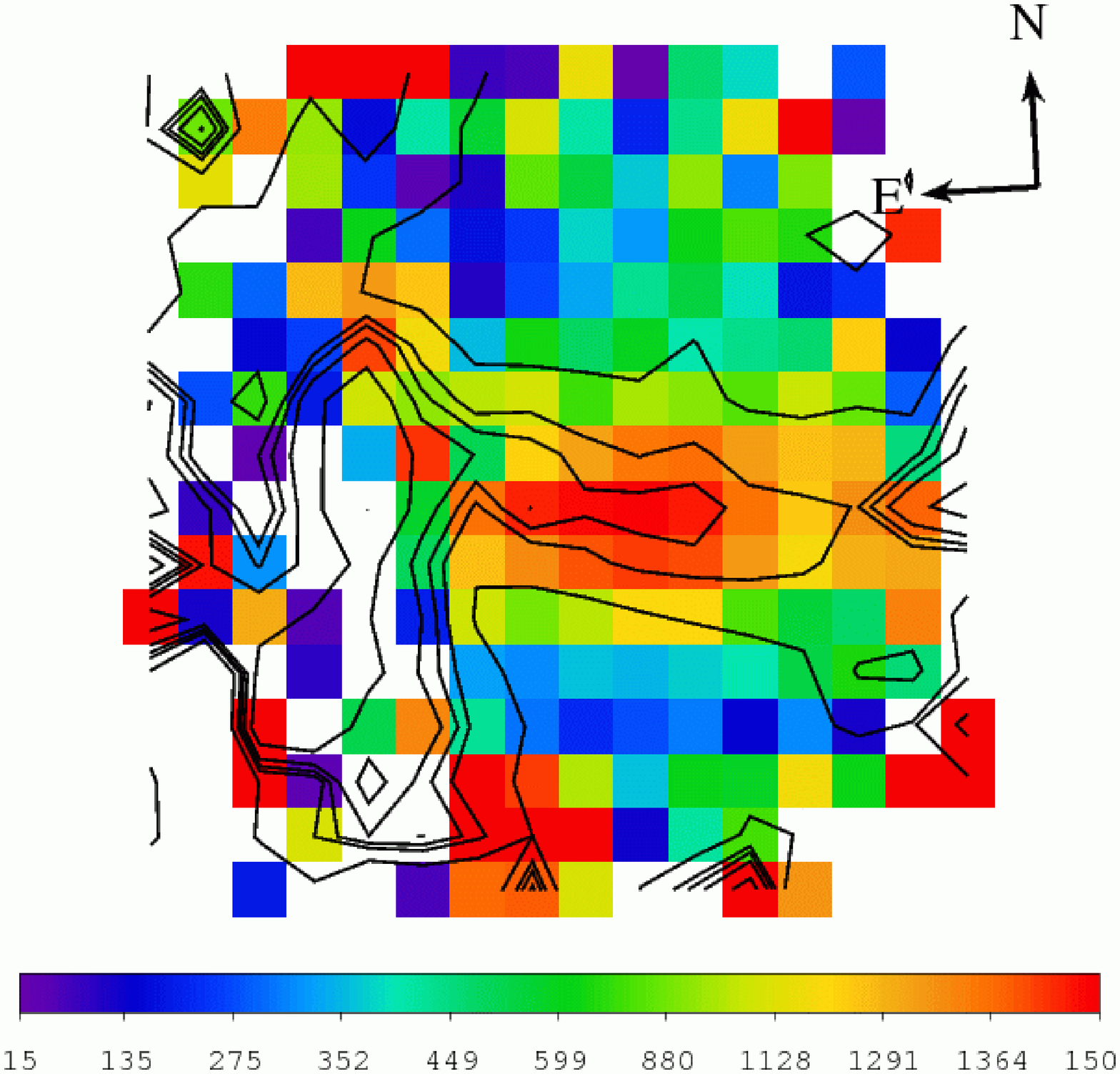}
\hspace{0.2cm}
 \includegraphics[width=4cm]{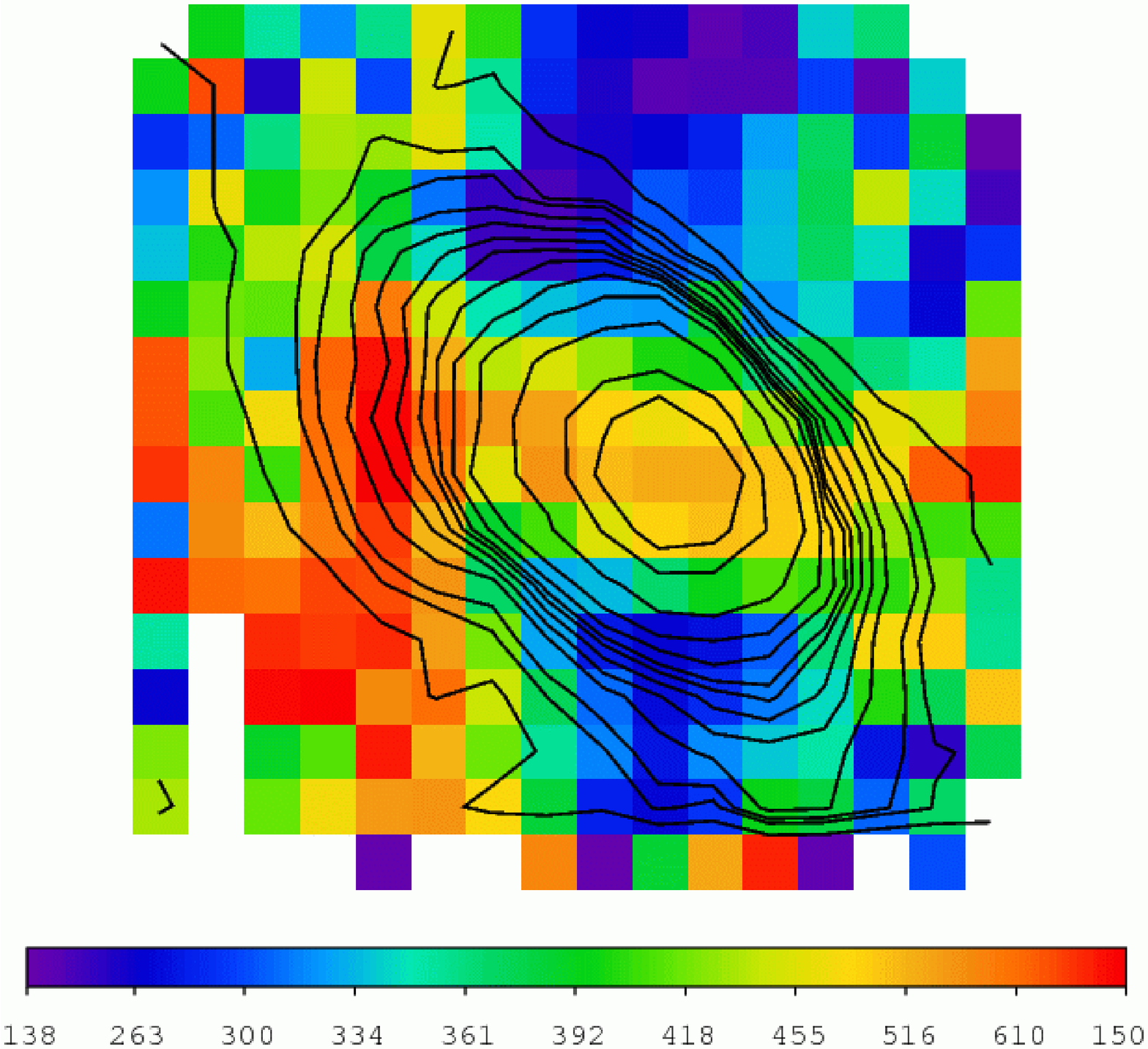}
\caption{The maps of electron density with the H$\alpha$ FWHM contours overlaid (left) and of H$\alpha$ FWHM with continuum contours overlaid (right).}
\label{ne_fwhm}
\end{figure}

\begin{figure}
\centering
 \includegraphics[width=9cm]{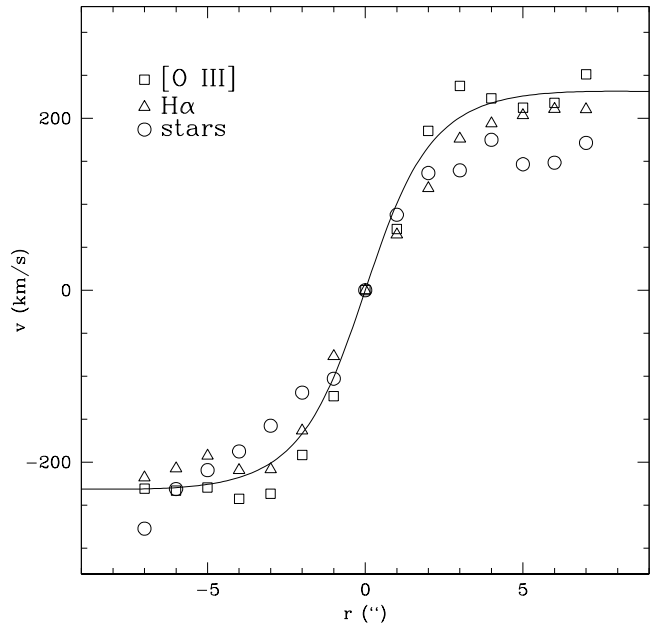}
\caption{Rotation curves for gas and stars. The solid line is the best-fitting of the gaseous component.
The median errors are 12 km s$^{-1}$ for [O\,{\sc iii}] and 19 km s$^{-1}$ for H$\alpha$. The spatial resolution is 1 arcsec = 516 pc.}
\label{rot_curve}
\end{figure}

For the main emission lines, we also built the 2D maps of velocity and FWHM (see in Fig. \ref{v600} two examples for [O\,{\sc iii}] and H$\alpha$). 
The measured velocities were corrected for the systemic velocity, considering the value obtained from the nuclear spectrum, that is the spectrum with the highest flux in the continuum map. 
For each spectrum the mean instrumental FWHM, obtained fitting few lines of the comparison lamp, was subtracted quadratically from the measured values, in order to take into account differences of the instrumental resolution in the field. 
The maps of the velocity field show values ranging from $-$180 km s$^{-1}$ to $+$150 km s$^{-1}$ for [O\,{\sc iii}] and from $-$220 km s$^{-1}$ to $+$180 km s$^{-1}$ for H$\alpha$. 
The maps of the FWHM, after the correction for instrumental width, show values between 200 km s$^{-1}$ and 800 km s$^{-1}$ for both [O\,{\sc iii}] and H$\alpha$ lines. 
The velocity maps show distortions East of the nucleus, especially in case of H$\alpha$, [N\,{\sc ii}] $\lambda$6548,6584 and [S\,{\sc ii}] $\lambda$6717,6731 lines, while the [O\,{\sc iii}] $\lambda$5007 line shows a more regular pattern.
In particular, the H$\alpha$ velocity map in the eastern region has velocity values between 130 km s$^{-1}$ and 200 km s$^{-1}$. 
This zone appears to be located towards the interaction region. Its high FWHM values (300--600 km s$^{-1}$) could be an effect of the gravitational interaction between NGC 7212 and its companion galaxy.
However, in the {\em HST} image and in the SAO broad-band image and color, this region shows no particular structure and has a low surface brightness ($10^{-16}$-$10^{-15}$ erg cm$^{-2}$ s$^{-1}$).  
In the [O\,{\sc iii}], [O\,{\sc i}], [N\,{\sc ii}], H$\alpha$ and [S\,{\sc ii}] FWHM maps it is clearly visible an elongated structure, with higher values, up to 700--800 km s$^{-1}$, oriented at about 90\degr\ with respect to the major axis of the emission. 
The FWHM of [O\,{\sc iii}] published by \citet{1990A&A...238...15D} ranges from 320 to 385 km s$^{-1}$. This disagreement is likely due to the orientations of their slit which do not correspond to this elongated structure. 
In any case, we observed values between 500 and 600 km s$^{-1}$ in the nuclear region.
This feature has been already observed in other Seyfert galaxies, like Mrk 3 \citep{DiMille2007}, Mrk 34, Mrk 1066, Mrk 348, Mrk 1, NGC 2992, NGC 5728 \citep{2009A&A...500.1287S}. The explanation of this effect is still under debate. This structure is also found when we fit the high resolution [O\,{\sc iii}] emission line profile with two separate components. 
The minor axis of the stellar velocity field is aligned with the photometric minor axis, on the contrary the gas velocity field is rotated with respect to the stellar one (Fig. \ref{v_gas_star}). 
From the analysis of the low resolution data we can see that stars and gas are not aligned, the [O\,{\sc iii}] emission is elongated along a direction orthogonal to the major axis of the galaxy. This could imply that the gas is not co-planar with the stars. 
It is interesting to comment that the jet-like structure, studied by \citet{1995ApJ...440..578T}, extending up to about 10 arcsec and oriented at PA $\sim$ $-$10\degr\ and showing redshifted velocity of 180 km s$^{-1}$, is not consistent with material ejected by the nuclear source, but instead it is rotating ionized gas as clearly visible from our [O\,{\sc iii}] velocity map. 
The density map is aligned as the FWHM maps. 
By comparing the high density ($N\rm _e = 700$--1400 cm$^{-3}$) regions in the map to the FWHM maps (see Fig. \ref{ne_fwhm}), we found high values of FWHM ($\sim$ 500 km s$^{-1}$) in regions showing high density. These are hints of compressed gas, or streaming of gas, and can be consistent with shocks.
The North-Eastern region with higher FWHM and a peculiarity in the velocity field is nearby the region of the merge between the galaxies and not connected to the ionization cone visible in the [O\,{\sc iii}]/H$\beta$ map. All the previous peculiarities could be caused by this interaction (see Fig. \ref{ne_fwhm}). 
From the analysis of the emission line profiles it is possible to point out the presence of asymmetries that can be due to additional and non-rotational kinematical components.
With the low resolution data, we were not able to identify and analyse multiple components in the spectra line profiles, although some asymmetries in the line profile for the higher S/N spectra were observed and, as a consequence, the fitting was more difficult to be performed.
From the higher resolution IFU data, initially we fitted [O\,{\sc iii}] and H$\alpha$ emission lines with a single component and we built the 2D maps of flux, velocity and FWHM. 
We used these maps as input for {\sc rotcur} \citep{1989A&A...223...47B}, the task of the {\sc gipsy} (Groningen Image Processing System) software, to model the 2D map and to obtain the deprojected values of the velocity in order to build the rotation curve. This task uses tilted-ring models: it models the velocity map using concentric rings and taking into account the inclination on the line of sight. Before starting, we needed to calculate the inclination angle (i) of the gas or stellar emission. 
We obtained i by matching the distribution of [O\,{\sc iii}] and H$\alpha$ emission with an ellipse. 
From this ellipse we also got the PA of the emission.
We corrected the velocity maps for the systemic velocity, using the values obtained in the central spectrum, and finally we could run {\sc rotcur}. We fixed the following input parameters: inclination (i = 40\degr), which was the same for the [O\,{\sc iii}] and the H$\alpha$ maps, the systemic velocity (0.0 km s$^{-1}$), the expansion velocity (0.0 km s$^{-1}$) and the coordinates (\emph{x}, \emph{y}) of the center of the galaxy. 
The other input parameters were the radii of the concentric rings in arcsec, the first guess for the rotational velocity of each ring and the position angle, and we let them free. 
To begin, we applied this task to the whole 2D velocity map, fitting both the approaching and the receding side at the same time. This output was used as input for the task VELFI which reconstructs the model of the observed velocity field, then we checked if the model could well reproduce the velocity map. 
On the other hand, to get more accurate values of the deprojected velocity in order to build the velocity curve, the velocity maps were fitted in two steps: first the receding and then the approaching side. 
We compared the gas rotation curves with the stellar rotation curve, obtained applying {\sc rotcur} to the stellar velocity map (i = 54\degr, PA$\sim$45\degr).
We found that the kinematical behaviour and the velocity values are similar for stars and gas. This implies that the kinematics of the ENLR gas and of the stars are dominated by the same gravitational potential (see Fig. \ref{rot_curve}). 
However, looking at the 2D maps we have already seen that the kinematical axis of stars is inclined of about 30\degr\ with respect to that of the gas (Fig. \ref{v_gas_star}). This could be a projection effect: gas and stars could be distributed in a different way and in a different plane. Stars are in agreement with the continuum emission, while gas velocity maps are oriented according to the gas emission.
We estimated the errors in velocity by measuring the position of some nightsky lines and assuming that the main contribution to the errors is due to wavelength calibration errors. 
We found median values of 12 km s$^{-1}$ for [O\,{\sc iii}] and 19 km s$^{-1}$ for H$\alpha$. No error is reported in the output of {\sc starlight} for stellar kinematics. 
In principle, we should take into account the S/N ratio of the emission lines \citep{1999A&A...342..671C}: when the S/N is high, the error is due only to errors in calibration, instead when the S/N is low, the error is due to the badly fitted line profile.
In our case, the S/N of these emission lines is high in the whole field. 
The errors may be larger in presence of asymmetries. 
For the high resolution data we also have the error due to the fitting of the two components. 
By assuming that gas and stars are on circular orbits in a plane, we can fit the rotation curves with the following formula \citep{1991ApJ...373..369B}:
\begin{equation}
v_c(r)=\frac{A r}{( r^2+ c_0^2)^{p/2}}
\end{equation}
where \emph{A}, $c_0$, and \emph{p} are parameters and \emph{r} is the radius in arcsec. The \emph{p} parameter is between 1 and 3/2.
We fit the gas rotation curve, using both [O\,{\sc iii}] and H$\alpha$ (see Fig. \ref{rot_curve}) with $A=300$ km s$^{-1}$, $c_0=2.5$ arcsec and $p=1.1$. The fitting of the stellar and gas rotation curve show a good agreement. 
By assuming a spherical potential, we can infer the mass distribution $M(r)$ using the Virial Theorem. Considering as maximum extension $r=7$ arcsec (corresponding to 3.6 kpc). Within this radius, we found $M=4.45\times10^{10}$ $\rm M_{\odot}$. 

\begin{figure}
 \centering
\includegraphics[width=4.1cm]{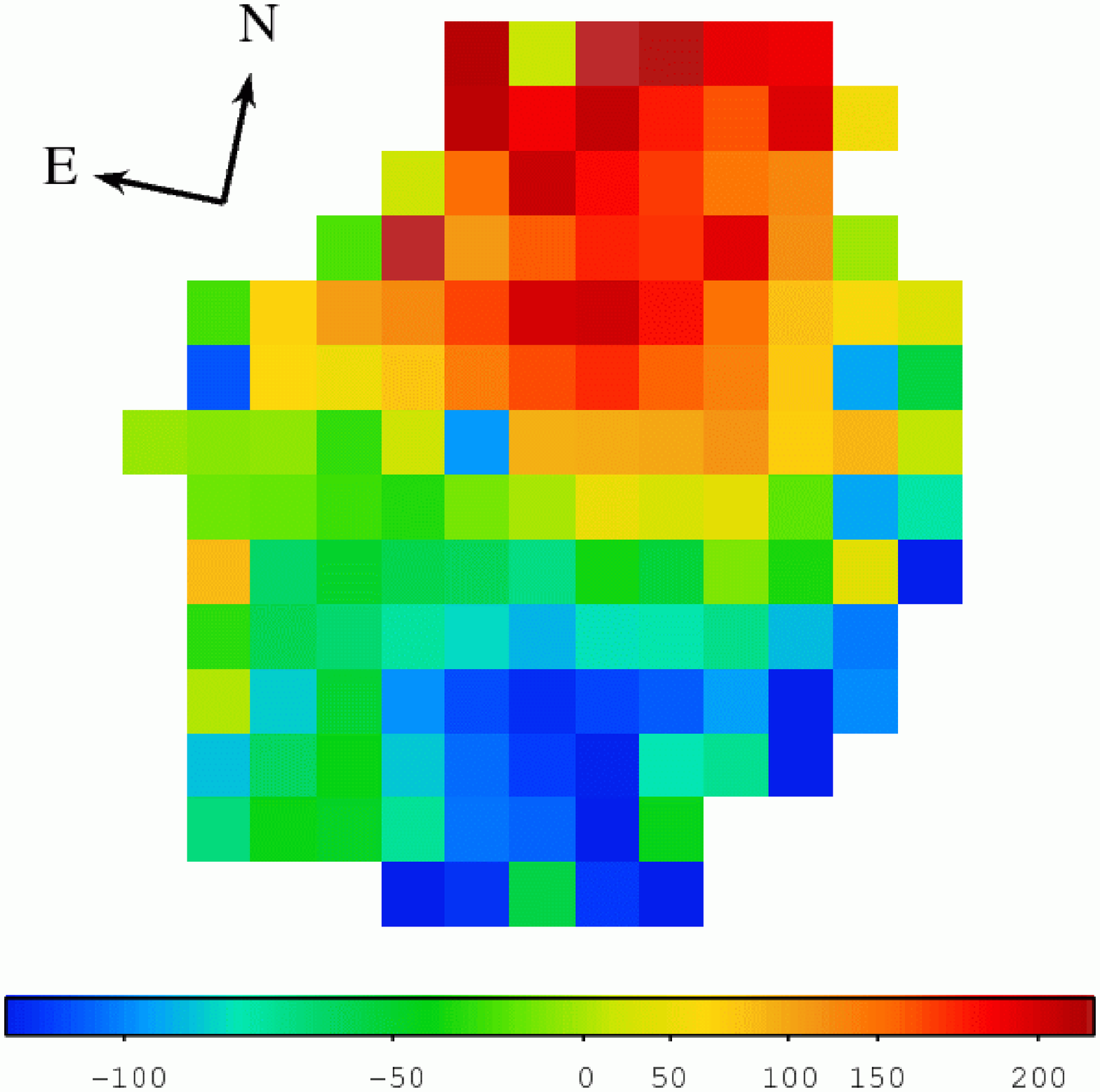}
\includegraphics[width=4.1cm]{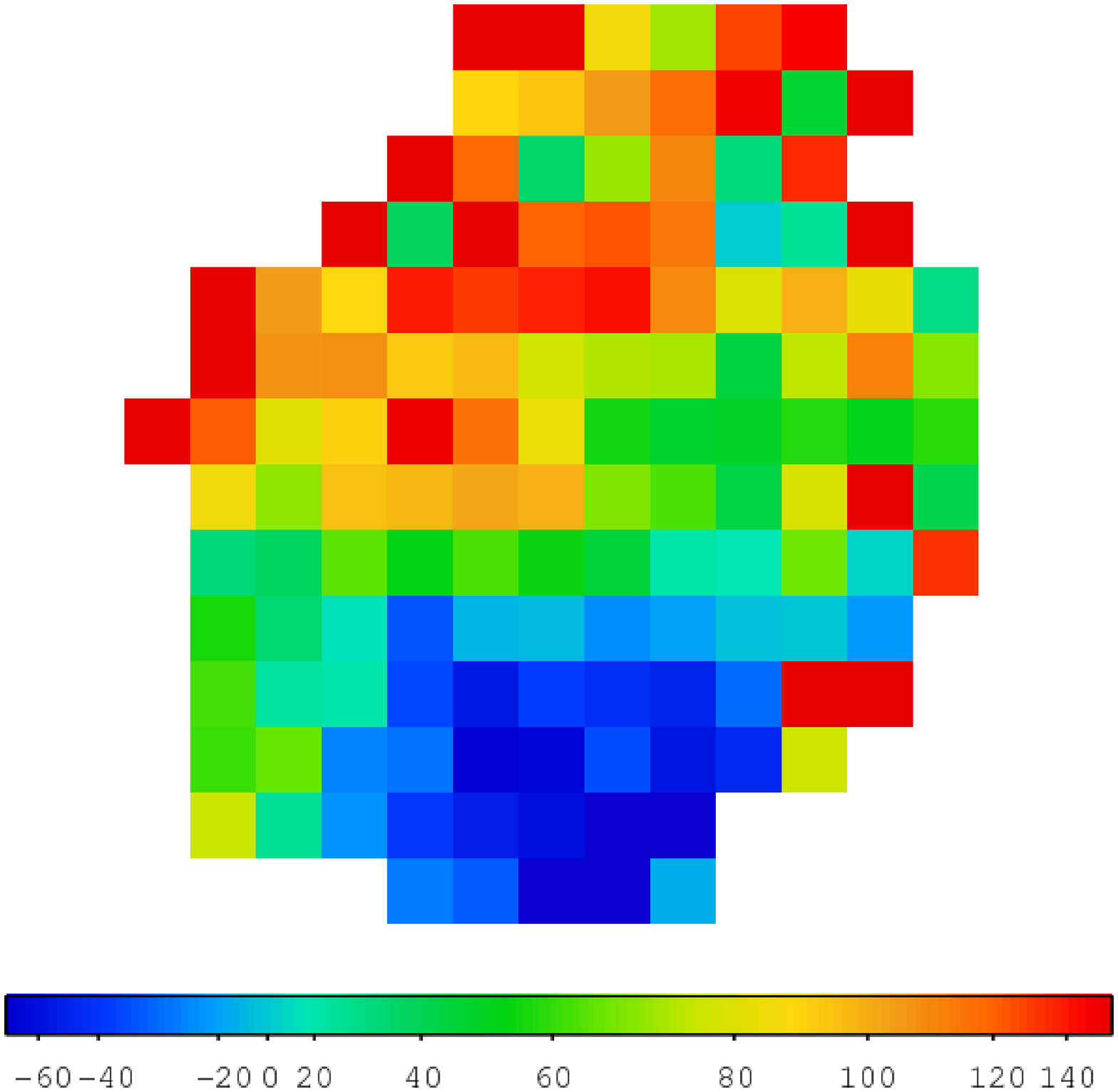}\\
\includegraphics[width=4.1cm]{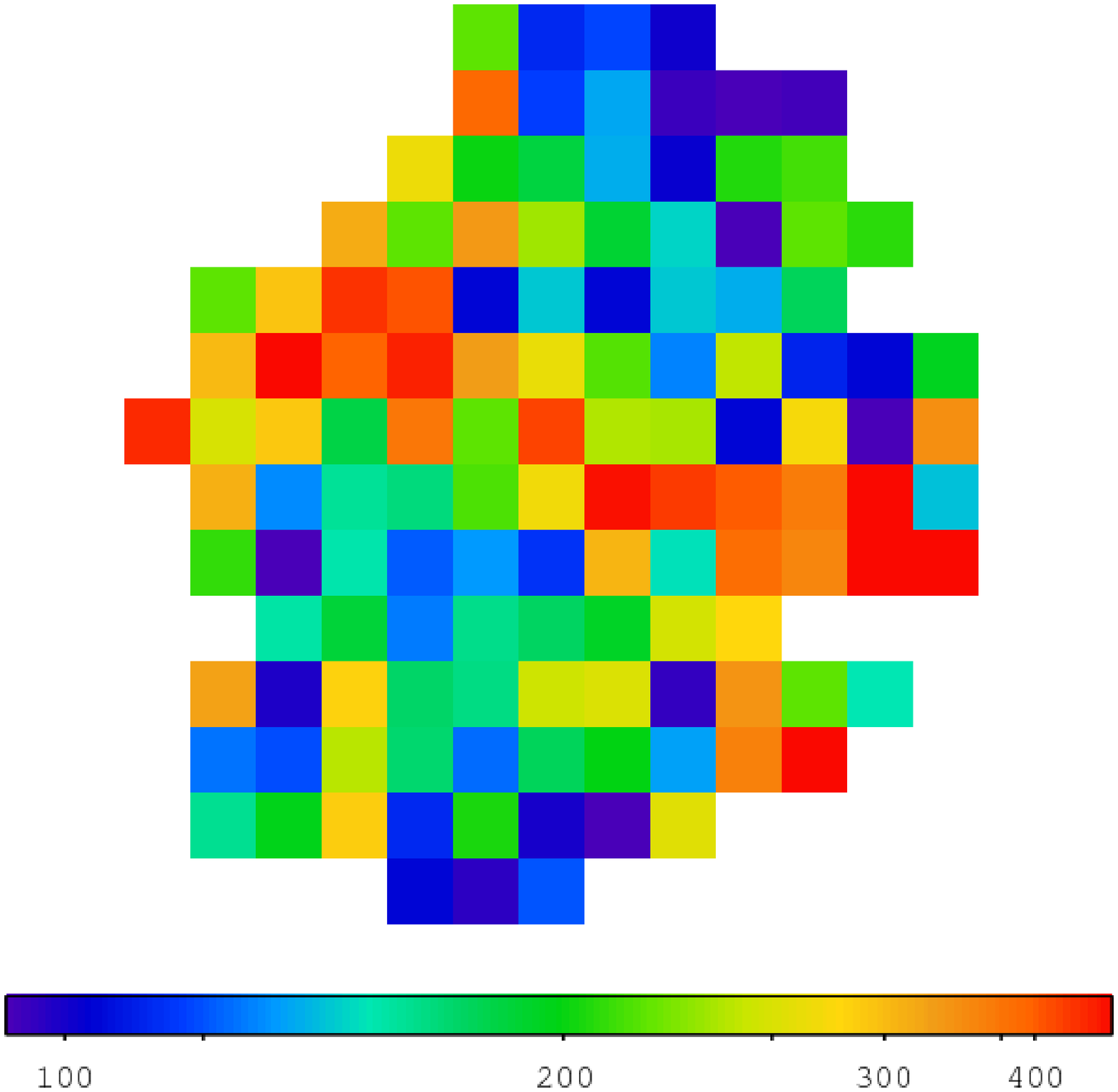}
\includegraphics[width=4.1cm]{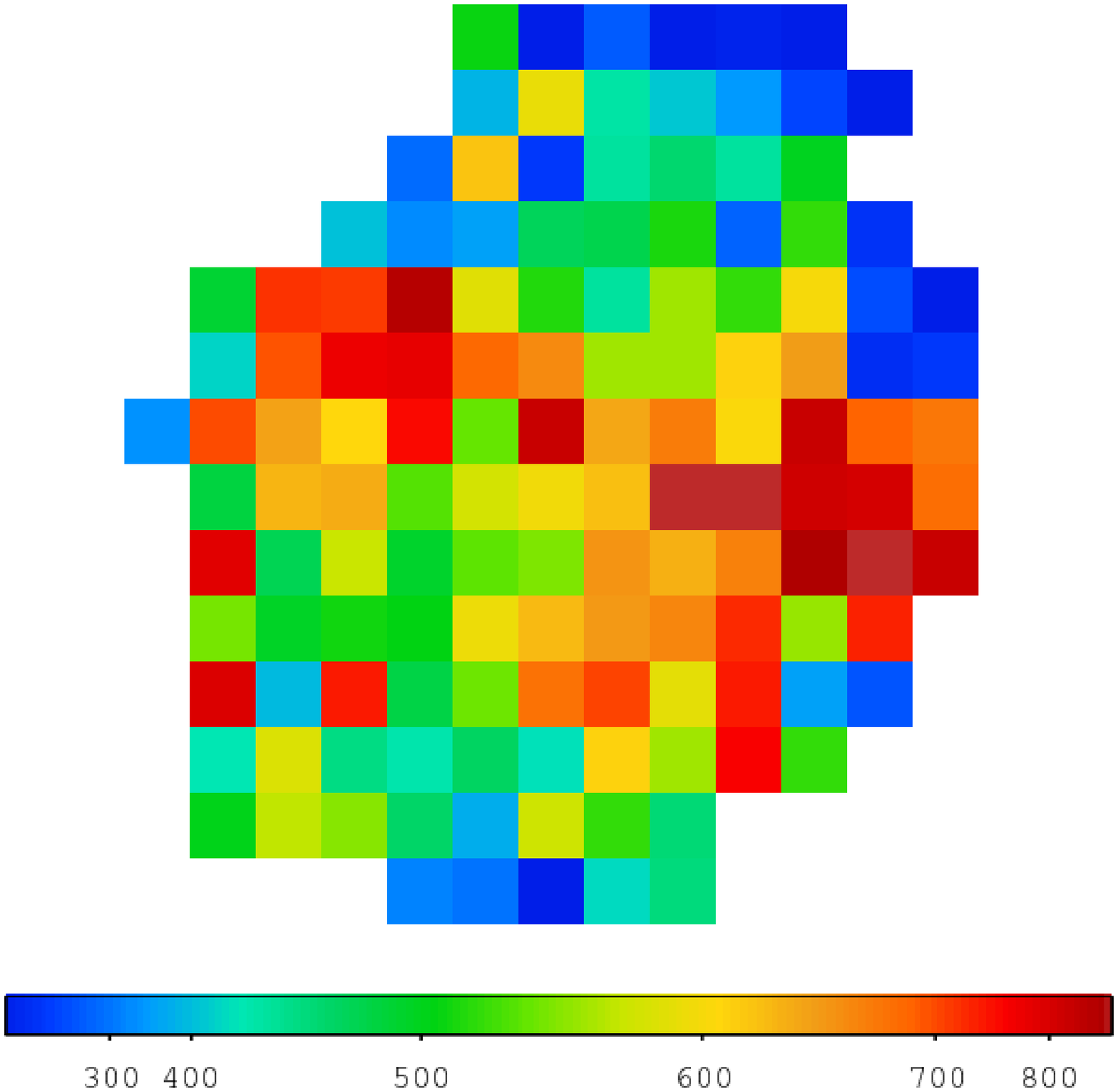}
\caption{Results of the multi-Gaussian fitting. Top:[O\,{\sc iii}] velocity map obtained for the the main component (left) and the broader one (right).
Bottom: FWHM maps for the main component (left) and the broader one (right).}
\label{v_comp}
\end{figure}

In order to perform a more accurate analysis of the kinematical behaviour of the ionized gas, we focused on the [O\,{\sc iii}] $\lambda$5007 line of the higher resolution MPFS spectra. 
By using a multiple Gaussian fitting, we identified two kinematical components, a narrow and a broad one, which suggest a complicated kinematic structure of the NGC 7212 ENLR. 
We decided to apply the {\sc midas} package {\sc xa}lice, which is more effective than {\sc pan} in this case, and we obtained the flux, velocity and FWHM maps for both narrow and broad components (see Fig. \ref{v_comp}). 
The two components have different velocities and different inclinations of the minor axis. 
For the narrow component we found values ranging between $-$120 km s$^{-1}$ and $+$170 km s$^{-1}$ and FWHM values up to 350--400 km s$^{-1}$.
The broader component shows lower velocity values, from $-$80 km s$^{-1}$ to $+$120 km s$^{-1}$ and FWHM up to 700--750 km s$^{-1}$. 
We estimated PA = 130\degr\ for the stellar map, PA = 97\degr\ for the narrow [O\,{\sc iii}] component and PA = 140\degr\ for the broad one. 
However, they are not aligned with the stellar velocity, as well.  
The high turbulence region observed in the single-fitting [O\,{\sc iii}] map and oriented as the minor axis of the gas emission is still present in both components.

\begin{figure}
\includegraphics[width=9cm]{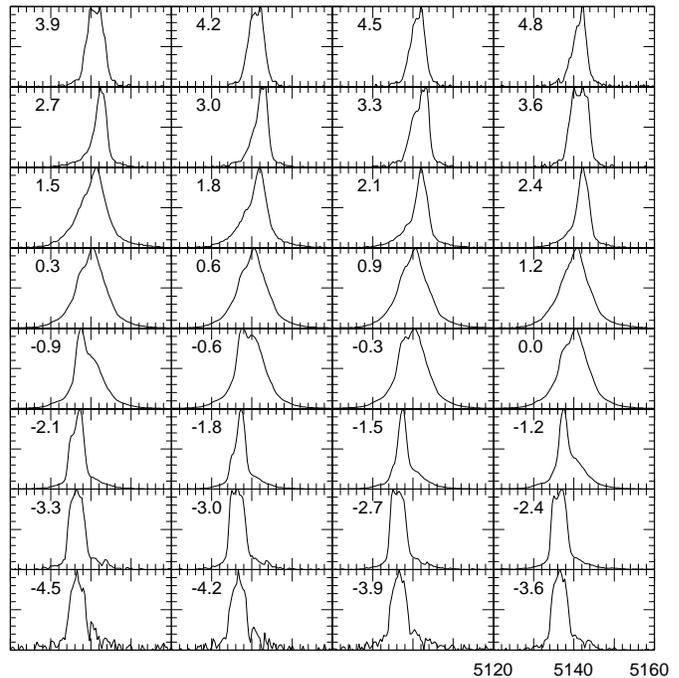}
\caption{The [O\,{\sc iii}] emission line profiles at different distances (in arcsec) from the nucleus, extracted from the echelle spectrum.}
\label{o3_profili}
\end{figure}

\begin{figure}
\includegraphics[width=8cm]{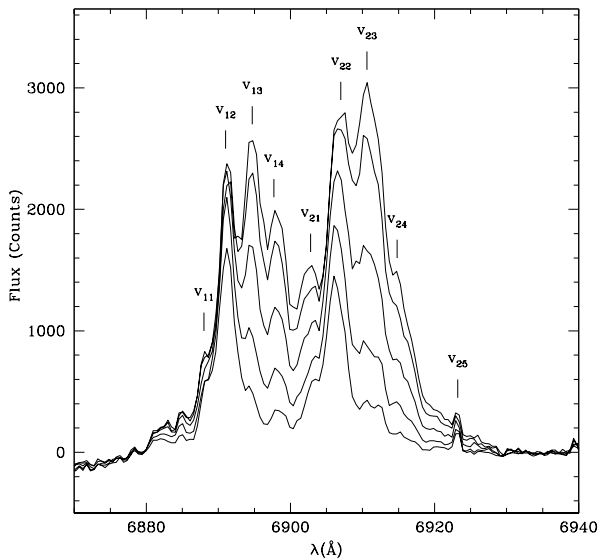}
\caption{[S\,{\sc ii}] emission line profiles at different distances from the nucleus. The peaks corresponding to the various components are consistent.}
\label{sii_echelle_comp}
\end{figure}

Having found hints of multiple components in [O\,{\sc iii}] profiles, we decided to explore the gas kinematics, by analysing a high resolution echelle spectrum (R$\sim$8000, instrumental FWHM$\sim$0.85 \AA) obtained with the slit oriented along the ionization cones.
By simply looking at the spectrum it is clear that the emission is due to various sub-components at different velocities.
We can see from Fig. \ref{o3_profili} that the emission line profile of [O\,{\sc iii}] is strongly variable and depending on the spatial position. 
Notwithstanding high spectral resolution, the components of the emission line profiles  of the high ionization gas are still broad, therefore we decided to analyse the lower ionization gas, using the [S\,{\sc ii}] doublet, in which the emission lines are narrower and it is easier to identify and distinguish the different components.
We found at least four components at different velocities (see Fig. \ref{sii_echelle_comp}): $v_{11}\sim -300$ km s$^{-1}$, $v_{12}\sim -150$ km s$^{-1}$, v$_{13}=0$ km s$ ^{-1}$ and $v_{14}\sim 170$ km s$^{-1}$, for the [S\,{\sc ii}] $\lambda$ 6716. These components correspond to the $v_{21}$, $v_{22}$, $v_{23}$ and $v_{24}$ of the [S\,{\sc ii}] $\lambda$6731.
The FWHM of the $v_{13}$ component of [S\,{\sc ii}] $\lambda$6716 is about 200 km s$^{-1}$. 
Maybe the lower velocity components are blended in the broad [O\,{\sc iii}] emission line profiles and are not detectable in lower resolution data. 
We found an higher velocity component for the [S\,{\sc ii}] $\lambda$6731 at v$_{25}\sim 450$ km s$^{-1}$, which is not visible in the [S\,{\sc ii}] $\lambda$6716 profile, because it is weak and it falls inside the [S\,{\sc ii}] $\lambda$6731 profile. 
It is worth noting that the ratio between the emission lines components changes with the distance from the nucleus, suggesting different physical conditions of the ionized gas. 
In fact, in the {\em HST} image of NGC 7212 it is clear that the ENLR is made of filaments, likely regions at different velocities, which are probably characterized by different physical parameters. 

\section[]{Ionization cone}\label{ioniz_cones}
\begin{figure*}
\includegraphics[width=5.3cm]{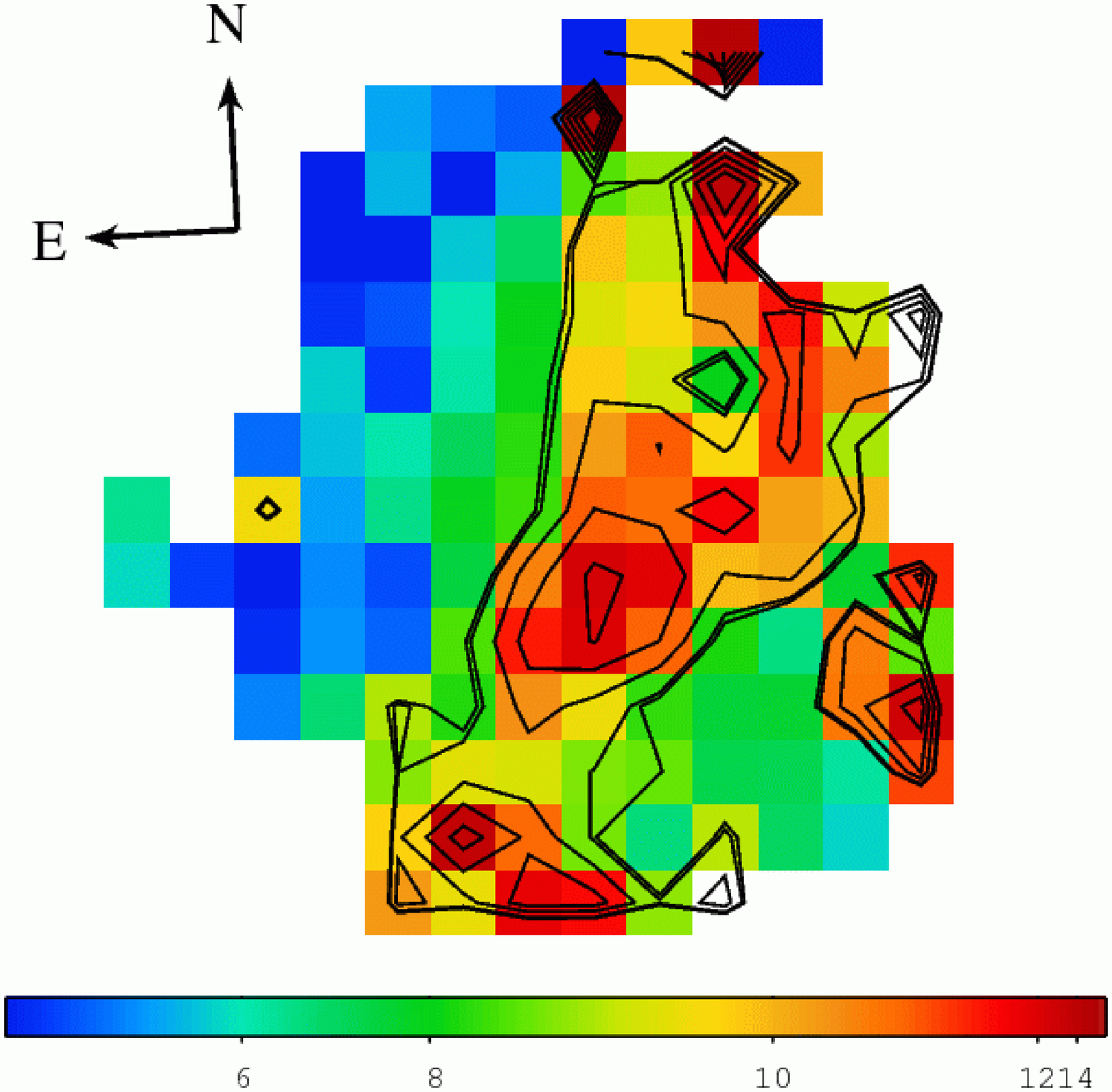}
\includegraphics[width=5.3cm]{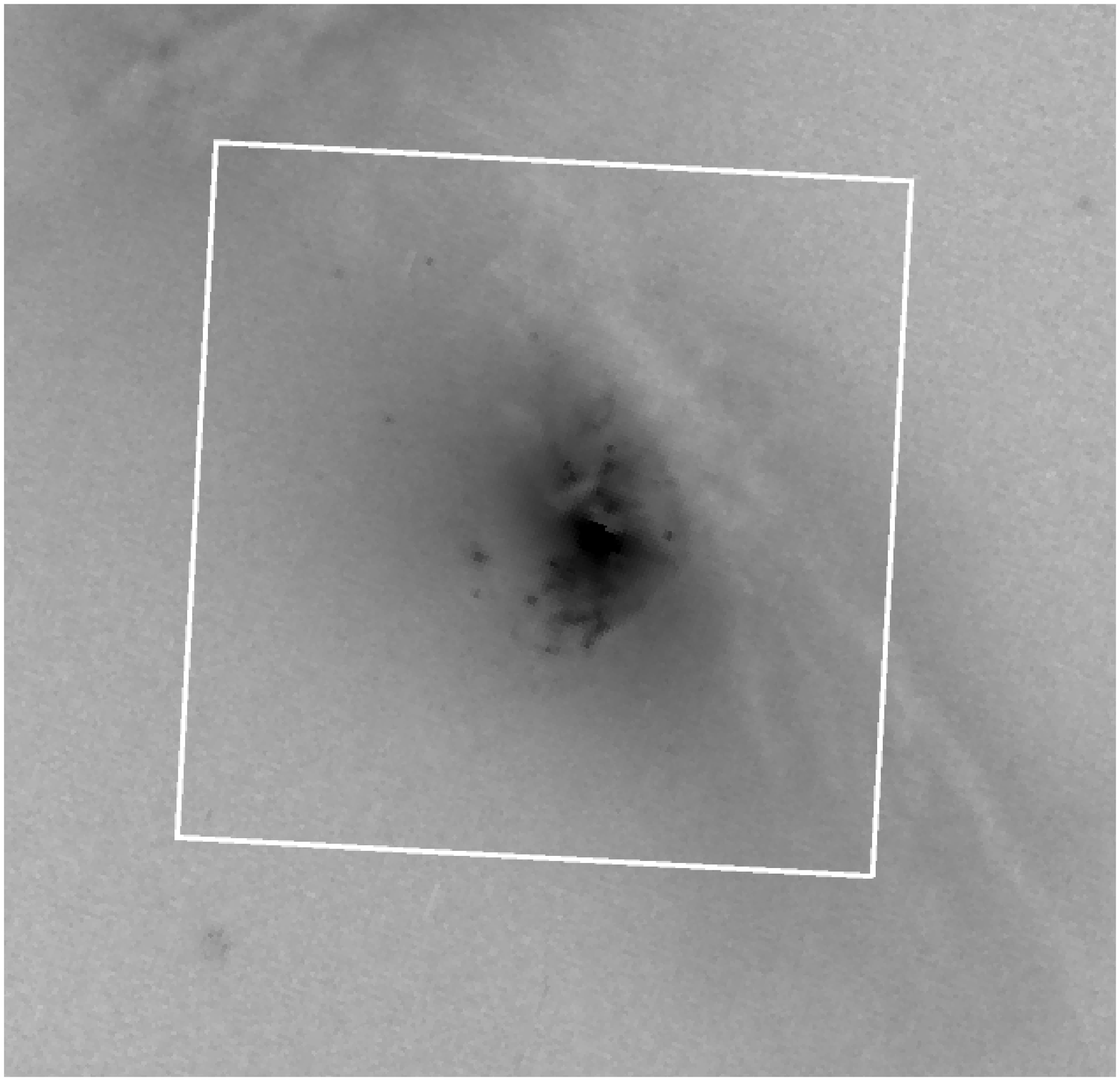}
\includegraphics[width=5.3cm]{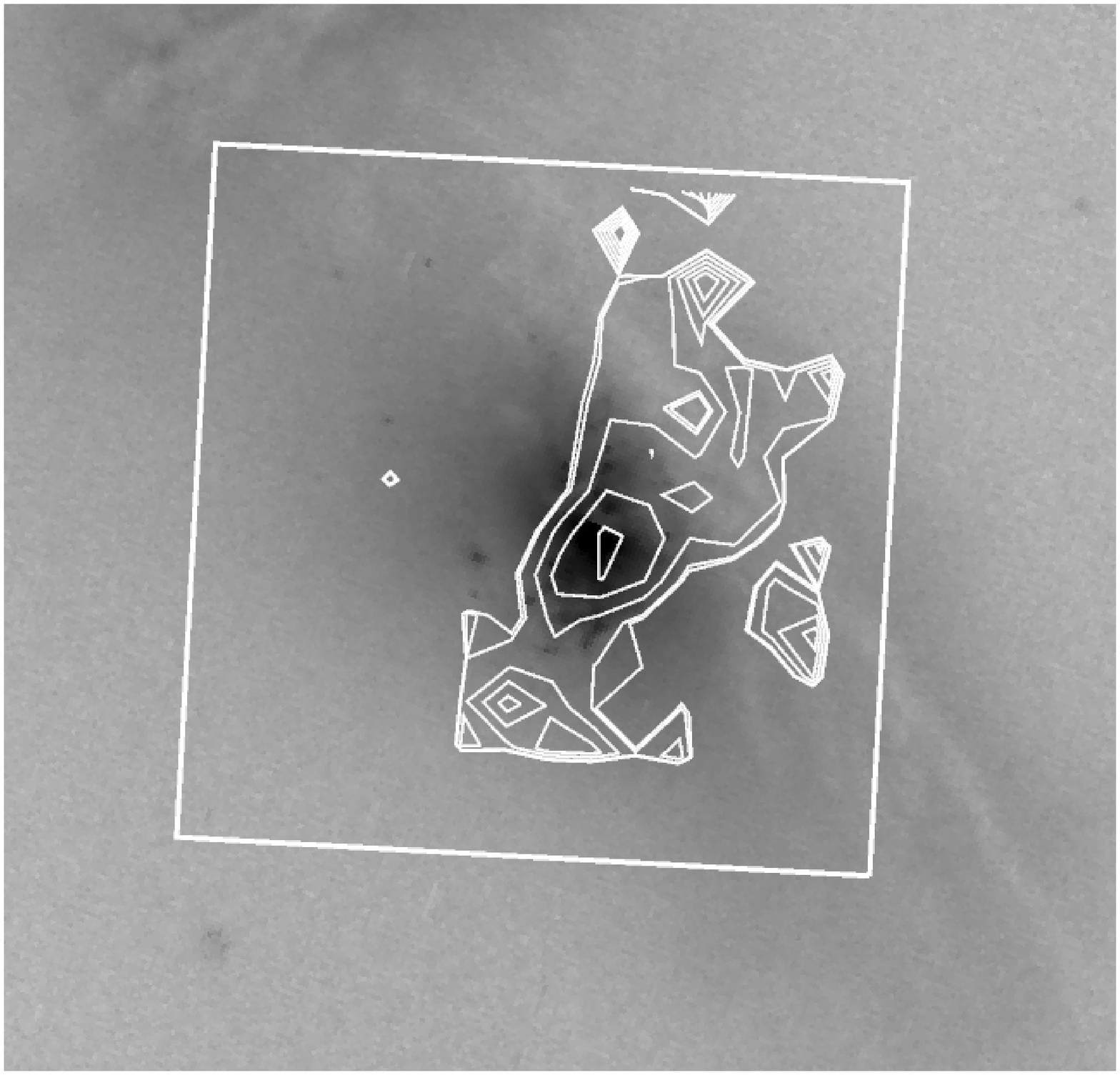}
\caption{The ionization map with the cone contours overlaid (left), the WFPC2 image of NGC 7212 (middle) and the same with the cone contours overlaid (right).}
\label{conoHST}
\end{figure*}

From the ionization map we have detected an elongated and extended highly ionized structure with high values (10--14) of [O\,{\sc iii}]/H$\beta$, up to 4 kpc from the nucleus, pointing out for the first time the presence of extended ionization cones in NGC 7212 (see Fig. \ref{conoHST}). 
In our map, this structure is clearly visible both North-West and South-East of the nucleus, while in \citet{1998ApJ...502..199F} only the south-eastern side is detected, probably because of the dust lane well visible in our $B-R$ image. 
In fact, they calculated the [O\,{\sc iii}]/H$\alpha$ ratio without any reddening correction. The same side was detected by \citet{2003ApJS..148..327S} with [O\,{\sc iii}] {\em HST} images. 
The lower limit of their measured flux is $\sim 6.5 \times 10^{-14}$ erg s$^{-1}$ cm$^{-2}$ arcsec$^{-2}$, while our detection limit is about $2\times 10^{-16}$ erg s$^{-1}$ cm$^{-2}$ arcsec$^{-2}$. 
\citet{1990A&A...238...15D} measured [O\,{\sc iii}]4959+5007/H$\beta$ $\sim$19 in the nucleus of NGC 7212. In their spectra this ratio ranges from 5 to 28. 
The reddening corrected value from \citet{2006A&A...456..953B} is $\sim16$ in the nucleus and varies between 6 and 17 in the central regions. 
Taking into account the [O\,{\sc iii}] $\lambda$4959 contribution, we found [O\,{\sc iii}]/H$\beta = 16$ in the nucleus, values ranging between 12 and 19 at about 7 arcsec North-West of the nucleus and between 12 and 16 at about 5.5 arcsec South-East of the nucleus. 
We selected the spaxels of the cone considering only those showing [O\,{\sc iii}]/H$\beta >8$. 
The aperture angle of the cone is $\sim$70\degr. As we have already mentioned, this angle could be larger to account for the regions located far from the nucleus in a direction orthogonal to the cones, but showing AGN-like ionization. 
The cone has a total size of 12 arcsec corresponding to $\sim$6 kpc and it is oriented at PA = 150\degr. 

\begin{figure}
 \centering
\includegraphics[width=8cm]{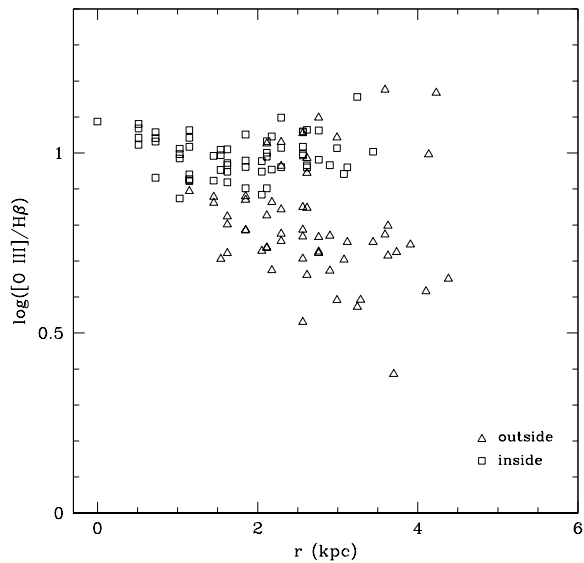}
\caption{The [O\,{\sc iii}]/H$\beta$ ratio as a function of the distance from the nucleus. The squares (triangles) are the spaxels inside (outside) the cone.}
\label{O3Hb_dist}
\end{figure}

The FWHM of the ionized gas inside the cone is around 200--300 km s$^{-1}$. The density is $<10^3$ cm$^{-3}$, without a radial trend, contrary to what found by \citet{2010A&A...516A...9D} in NGC 5252. These authors derived a $r^{-2}$ relation from the constant [O\,{\sc iii}]/(0.5--2 keV) flux ratio, which suggests a radial-independence of the ionization parameter.
We observed a gradient of temperature, with higher values in the northern region.
We plotted the [O\,{\sc iii}]/H$\beta$ ratio versus the distance from the nucleus, dividing the spaxels inside and outside the cone (Fig. \ref{O3Hb_dist}).
As expected, we found that outside the cone the ratio decreases with the distance from the nucleus, while inside the cone the ratio is high even far from the nucleus (up to 4 kpc). 
In order to understand if the AGN can account for such a high level of ionization, we made some energy balance considerations.
Firstly, we calculated the L(H$\alpha$) from the reddening corrected H$\alpha$ flux.
Then the number of ionizing photons needed to produce such luminosities,
$ N_{ph} = 7.3 \times 10^{11}  {\rm L(H}\alpha)$ ph s$^{-1}$ \citep{1998ARA&A..36..189K}, was estimated and compared with the number of photons emitted by the active nucleus, diluted by the covering factor $\Omega/4\pi$, where $\Omega$ is the solid angle of a region as seen from the nucleus. We did not make any hypothesis about the spatial distribution of the gas within each region.
By assuming that the nuclear source is an isotropic emitter of radiation absorbed along the line of sight, we estimated the nuclear ionizing photons by averaging the calculated number of photons ionizing the regions surrounding the nucleus and located within the cones, after having removed the effect of the covering factor. 
We found $N^{nuc}_{ph} = 9.4\times10^{53}$ s$^{-1}$. 
The median value of the ratio between the observed value and the diluted one is 0.84, indicating that the active nucleus is the dominant ionizing source of the regions within the cone. 

Estimating the mass of gas can give an idea of the origin of the gas, in fact a large amount of gas can be explained with the acquisition of external material resulting from a merger or an interaction event. 
The mass of the gas in the ionization cone was estimated following two methods. The first one makes use of the H$\alpha$ luminosity, as explained by \citet{2009ApJ...699..638H}:
\begin{equation}
 M(L_{H\alpha}) = 2.97 \times 10^3 \left(\frac{100~{\rm cm^{-3}}}{N_e}\right) \left(\frac{L_{H\alpha}}{{\rm 10^{38}~erg~s^{-1}}}\right) {\rm M_\odot} 
\label{Ho_eq}
\end{equation}
the second one assumes a Galactic dust-to-gas ratio and makes use of the interstellar extinction \citet{2007ApJ...666..794F}. We used a modified version of their formula:
\begin{equation}
 M(A_V) = 1.5 \times 10^7 {\rm kpc^{-2} mag^{-1}} A_V (s \times \theta)^2 {\rm M_\odot} 
\end{equation}
where $s$ is the scale in kpc arcsec$^{-1}$ and $\theta$ is the size of a region in arcsec.
The results of the two methods are different: in particular, the second method gives larger values. 
The total values are $M(L_{H\alpha})=5.16 \times 10^6 M_\odot$ and $M(A_V)=3.16 \times 10^8 M_\odot$.
The reasons of this discrepancy could be a combination of two effects. First, the $N_e$ value used in Eq. \ref{Ho_eq} is generally based on the [S\,{\sc ii}]6716/[S\,{\sc ii}]6731 ratio, which fails in case of low electron density. Second, the flux of recombination lines depends on electron density squared. 
Therefore, the first method can cause an underestimate of the ENLR mass, which is likely constituted by a large fraction of low density gas.

\begin{figure}
 \centering
\includegraphics[width=7.5cm]{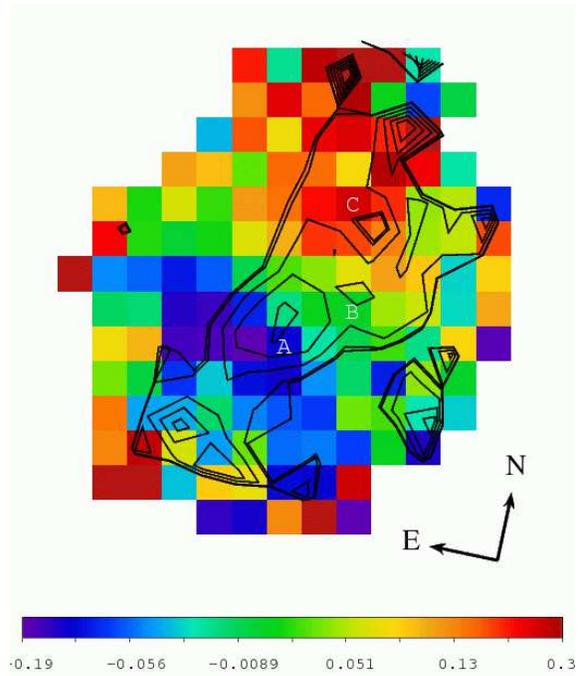}
\includegraphics[width=7.5cm]{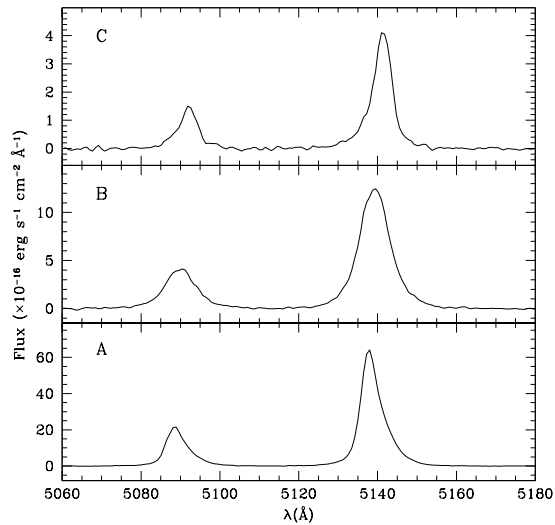}
\caption{Top: the asymmetry map with the cone contour overlaid.
Bottom: three examples of [O\,{\sc iii}] profiles in different regions of the FoV. Panel A shows a spectrum with asymmetry equals to $-$0.14, panel B shows a profile without asymmetry, panel C shows a spectrum with asymmetry=0.21.}
\label{asymm}
\end{figure}
We measured the asymmetry of the [O\,{\sc iii}] emission lines by applying the method published by \citet{1985MNRAS.213....1W} to the higher resolution spectra. 
We calculated the area under the emission line and the wavelength corresponding to the 10 per cent ($\lambda_{10}$), 50 per cent ($\lambda_{50}$) and 90 per cent ($\lambda_{90}$) of the emission line area. Then we obtained the asymmetry asym$=\frac{a-b}{a+b}$ where $a=\lambda_{50}-\lambda_{10}$ and $b=\lambda_{90}-\lambda_{50}$.  
In case of a blue wing in the line profile, the asymmetry is positive, while when a red wing is present, the asymmetry is negative.
We reconstructed the asymmetry map (Fig. \ref{asymm}) which shows two distinct regions, one with positive (C region, in Fig. \ref{asymm}), and the other with negative asymmetry (A region, blue colours in Fig. \ref{asymm}). These two regions match the ionization cone contours. 
We found [O\,{\sc iii}] emission line profiles with blue wings in the northern part of the ionization cone, while the southern part of the cone is characterized by [O\,{\sc iii}] with red wings. In the nucleus we found a symmetric profile. 
This suggests the presence of gas in radial motions, outflow or inflow, inside the cone. 
The asymmetry is not correspondent to the higher FWHM regions, that is orthogonal to the cone, therefore these high values of FWHM are not an effect of multiple kinematical components.

\section[]{Conclusions}
We studied the physical and kinematical properties of the circumnuclear gas in the nearby Seyfert 2 galaxy NGC 7212. We analysed this object by means of integral field and echelle spectra and broad-band images.
We pointed out for the first time the presence of an extended ionization cone in NGC 7212, with high values of [O\,{\sc iii}]/H$\beta$ (up to 12), at a large distance from the nucleus (up to 3.6 kpc).   
The cone is oriented NW-SE, at PA = 150\degr\ close to the photometric minor axis of the galaxy, with an opening angle of about 70\degr. The cone is more extended in the North (7 arcsec = 3.6 kpc) than in the South (5.5 arcsec = 2.8 kpc) direction, while in {\em HST} [O\,{\sc iii}] published images only the southern high ionization emission was detected, showing a structure made of clouds or filaments.
The evidence of dust located in the ENLR also supports this idea that the gas in the cone has a filamentary structure, and suggests that a potential source of error raises in the dereddening, because how the gas and dust are mixed in and between filaments is in fact unknown.
The mass of the ENLR was calculated by means of two different methods based on the H$\alpha$ luminosity and on the interstellar extinction, and it is likely between 5$\times10^6$ and 3$\times10^8$ M$_{\odot}$.
NGC 7212 is in an interacting triplet, with two galaxies in a clear on-going merger. We found high values of [N\,{\sc ii}]/H$\alpha$ and [S\,{\sc ii}]/H$\alpha$ towards the interaction region: suggesting a possible combination of ionization by the active nucleus and by shocks.
From the 2D velocity maps, we found kinematical decoupling between the stars and gas. The velocity fields are misaligned, with the stellar kinematical minor axis aligned with the photometric minor axis, while the gas kinematical axis is tilted 30\degr\ with respect to the stellar one.
We studied the asymmetry of the emission line profiles inside the ionization cone and we found [O\,{\sc iii}] with blue wings in the northern side of the cone, while the southern side of the cone is characterized by [O\,{\sc iii}] with red wings. In the nucleus the profiles are symmetric. This suggests the presence of gas in radial motions, which is confirmed also by the analysis of high resolution spectra. In fact, studying the echelle data, we found that the ionized gas is characterized by multiple kinematical components at different velocities. We found at least four components: at $-$300 km s$^{-1}$, $-$150 km s$^{-1}$, 170 km s$^{-1}$ and 450 km s$^{-1}$, with respect to the recession velocity. 
The ENLR gas metallicity was estimated by measuring the observed emission line ratios and comparing them with {\sc cloudy} models, obtaining indications of sub-solar values. 
All these properties support the idea of an external origin of the ENLR gas in NGC 7212, likely due to gravitational interaction effects in act in this triple system.

\section*{Acknowledgments}
We are greatful to the anonymous referee for useful comments and suggestions which improved the quality of the paper.

F.D.M. acknowledges the support of a Magellan Fellowship from Astronomy Australia Limited, and administered by the Australian Astronomical Observatory.

Australian access to the Magellan Telescopes was supported through the National Collaborative Research Infrastructure Strategy of the Australian Federal Government. 

This research has made use of the NASA/IPAC Extragalactic Database (NED) which is operated by the Jet Propulsion Laboratory, California Institute of Technology, under contract with the National Aeronautics and Space Administration. 

V.C. dedicates this paper to her husband Andrea.

\end{document}